\documentclass[10pt,letterpaper,notitlepage]{article}
\usepackage[utf8]{inputenc}
\usepackage{amsmath}
\usepackage{amsfonts}
\usepackage{amssymb}
\usepackage{graphicx}
\usepackage{cancel}
\usepackage{float}

\usepackage[ruled,vlined]{algorithm2e}

\usepackage[left=0.75in, right=0.75in, bottom=1.0in,top=0.75in]{geometry}

\usepackage{caption} 
\newcommand{\DOCTITLE}{Massively Parallel Transport Sweeps on Meshes with Cyclic Dependencies
}  

\newcommand{\beq}{\begin{equation*}
\begin{aligned}}
\newcommand{\eeq}{\end{aligned}
\end{equation*}}

\newcommand{\beqn}{\begin{equation}
	\begin{aligned}}
\newcommand{\eeqn}{\end{aligned}
	\end{equation}}  

\newcommand{\bea}{\begin{eqnarray}}
\newcommand{\eea}{\end{eqnarray}}
\newcommand{\be}{\begin{equation}}
\newcommand{\ee}{\end{equation}}
\newcommand{\beas}{\begin{eqnarray*}}
\newcommand{\eeas}{\end{eqnarray*}}
\newcommand{\bdm}{\begin{displaymath}}
\newcommand{\edm}{\end{displaymath}}
\newcommand{\bl}{\bss\begin{itemize}}
\newcommand{\el}{\vspace{-.5\baselineskip}\end{itemize}\ess}
\newcommand{\ben}{\bss\begin{enumerate}}
\newcommand{\een}{\vspace{-.5\baselineskip}\end{enumerate}\ess}
\newcommand{\bfg}{\begin{figure}}
\newcommand{\efg}{\end{figure}}
\newcommand{\bt}{\begin{table}}
\newcommand{\et}{\end{table}}
\newcommand{\bc}{\begin{center}}
\newcommand{\ec}{\end{center}}
\newcommand{\btb}{\begin{center}\begin{tabular}}
\newcommand{\etb}{\end{tabular}\end{center}}
\newcommand{\bss}{\begin{singlespace}}
\newcommand{\ess}{\end{singlespace}}
\newcommand{\bi}{\begin{itemize}}
\newcommand{\ei}{\end{itemize}}

\newcommand{\Omegabf}{\mathbf{\Omega}}
\newcommand{\position}{\mathbf{r}}

\usepackage[normalem]{ulem}

\begin{document}
\noindent
{\LARGE\textbf{\DOCTITLE}}
\newline
\newline
\newline
\noindent
{\Large Jan I.C. Vermaak${^1}$, Jean C. Ragusa$^{1,2}$, Jim E. Morel$^{1,2}$}
\newline
\noindent\rule{\textwidth}{1pt}
{\small $^1$Center for Large Scale Scientific Simulations, Texas A\&M Engineering Experiment Station, College Station, Texas, USA.}
\newline\noindent
{\small $^2$Nuclear Engineering Department, Texas A\&M University, College Station, Texas, USA.}
\newline
\newline
\textbf{Abstract:}\newline\noindent
When solving the first-order form of the linear Boltzmann equation, a common misconception is that the matrix-free computational method of ``sweeping the mesh", used in conjunction with the Discrete Ordinates method, is too complex or does not scale well enough to be implemented in modern high performance computing codes. This has led to considerable efforts in the development of matrix-based methods that are computationally expensive and is partly driven by the requirements placed on modern spatial discretizations. In particular, modern transport codes are required to support higher order elements, a concept that invariably adds a lot of complexity to sweeps because of the introduction of cyclic dependencies with curved mesh cells. In this article we will present a comprehensive implementation of sweeping, to a piecewise-linear DFEM spatial discretization with particular focus on handling cyclic dependencies and possible extensions to higher order spatial discretizations. These methods are implemented in a new C++ simulation framework called Chi-Tech ($\chi{-}Tech$). We present some typical simulation results with some performance aspects that one can expect during real world simulations, we also present a scaling study to $>$100k processes where Chi-Tech maintains greater than 80\% efficiency solving a total of 87.7 trillion angular flux unknowns for a 116 group simulation.

\section{Introduction}
The linear Boltzmann transport equation \cite{LewisMiller} is used as a mathematical model for describing many real-world physical phenomena such as neutral particle transport (neutrons, photons, etc.), radiative heat transfer, and the scattering of light off surfaces, among others. Using a suitable spatial discretization results in a linear system with a matrix that is logically block lower triangular which can be solved with a single forward solve referred to as a ``sweep". In practice however, the transport equation is coupled to other angles resulting in the need to solve this equation for the entire phase space including energy. Discretizing all the dimensions results in a large amount of unknowns that can benefit from parallel computing. To this end parallel sweeps have become an important method to solve the linear Boltzmann transport equation.

A popular method for parallel sweeping is the Koch, Baker and Alcouffe (KBA) method \cite{KBA} whereby a specific parallel partitioning of cells is used along with a topological sorting, for a given direction, of these cells and partitions. The partitioning referred to here is to divide the problem domain into a sub-grid of $P_x{\times}P_y$ orthogonal partitions. This partitioning method was extended to $P_x{\times}P_y{\times}P_z$ in 2013 by Adams et al. \cite{ProvablyOptimal0} where they also developed a concise performance model. They were able to show that by sweeping angles from all the octants simultaneously and by using an appropriate sweep scheduling algorithm, parallel transport sweeps can scale to millions of processes. The analysis of a data-driven sweep scheduling, by Bailey and Falgout in 2009 \cite{Analysis2009}, also allowed Adams to formulate the Depth-Of-Graph sweep scheduling algorithm which, apart from determining which sweep directions have priority to execute on a given partition, also addresses how tie breakers are handled. Adams et al. \cite{ProvablyOptimal0} implemented these algorithms in their C++ based Parallel Deterministic Transport (PDT) code, built on the STAPL parallel library \cite{STAPL}, and used piecewise linear Discontinuous Finite Element discretization on polygons and polyhedra as introduced by Stone and Adams in 2003 \cite{Stone} and implemented in PDT by Bailey in 2008 \cite{PolyFEM}\cite{Poly2FEM}. The piecewise linear discretization allowed PDT to use polygons and polyhedrons in transport simulations and created many capabilities to perform transport simulations on unstructured meshes.

PDT was also used in the recent paper by Adams \cite{ProvablyOptimal} to showcase unstructured mesh capability as opposed to the orthogonal meshes used in \cite{ProvablyOptimal0}. The results showed that, for the KBA style partitioning, the parallel performance was still optimal and demonstrated scaling up to $\sim$1.5 million processes. Adams also performed a study in 2015 \cite{ProvablyOptimalb} that demonstrated the use of non-contiguous partitions and found that partition over-loading (assigning partitions to processes that experience large amounts of idle-time), can improve parallel efficiency over their non-overloaded counterparts. This partitioning style however, needs to weighed against load balancing efforts such as those discussed by Ghaddar and Ragusa \cite{Ghaddar2017}. In the transport sweep studies by Adams et al. \cite{ProvablyOptimal0}\cite{ProvablyOptimal}, the investigations were performed in using optimal conditions, therefore, the performance models were developed for perfectly balanced meshes. 

Ghaddar et al. also studied load balancing in 2019 \cite{Ghaddar2019} with specific reference to non-KBA style partitionings. The types of partitions studied were still brick shaped but not orthogonal like that used in KBA-partitions. Ghaddar's study, using load balancing by dimension, proved that a perfectly balanced non-orthogonal partition does not ensure optimal parallel sweep efficiency. This effect has direct bearing on our unstructured mesh efforts and will form a critical portion of our future work.

In addition to the review of parallel sweeping in this introduction, we also review literature with regards to sweeping on meshes that exhibit cyclic dependencies. The KBA-style partitioning and its non-orthogonal variant normally implies the use of convex brick-shapes for the partition. Volumetric partitioning schemes such as those employed in the ParMETIS code \cite{ParMETIS} ultimately result in partitions with very degenerate polyhedral shapes which is of no concern to elliptic problems, but for sweep algorithms this results in re-entrant partition boundaries referred to in this paper as global cyclic dependencies. Although we will focus mainly on global cyclic dependencies in this paper we will also touch on local cyclic dependencies introduced by concave cell faces and opposing reflecting boundary conditions. Local cyclic dependencies are also inevitably created when using curvilinear coordinates. Haut et al. \cite{EdgeSplitting} used face/edge splitting to remove the cycles and Liu and Larsen \cite{CurvedGrids} used cell splitting to achieve the same. These methods had the effect of removing the cyclic dependency problem entirely. Another solution to the cyclic dependency is to defer these cyclic dependencies to old angular fluxes, in other words ``lagging" the cyclic dependency information to previous iterations. Both Haut et al. \cite{RemoveCycleLLNL} and Plimpton et al. \cite{RemoveCycles} devised such methods but where Plimpton et al. continued to use source iteration for their iterative scheme, Haut depicted a formulation which could also use GMRES. If GMRES is to be used, the cyclically dependent angular fluxes has to be included in the vector of unknowns which was not done by Haut. Wang and Ragusa \cite{RemoveCyclesYaqi} used the same formulation and indeed included their cyclically dependent unknowns in the vector of unknowns.
\newline
\newline
In this article we will introduce the C++ simulation platform called Chi-Tech. This code is a relatively new simulation platform that implements fully contained parallel sweep functionality, using the Message Passing Interface (MPI), with the literature discussed above as a basis. The code uses PETSc \cite{PETSc} for implementation of the parallel GMRES algorithm as well as the iterative methods used with Diffusion Synthetic Acceleration (with HYPRE's \cite{HYPRE} Boomer Algebraic Multi-Grid as preconditioner).
\newline
\newline
We start our discussions with some background on the application of the discrete ordinates method, particularly quadrature rules and iterative algorithms. We then discuss the application of the discontinuous Finite Element method on polygons and polyhedrons which lead into our discussions regarding transport sweeps. We then discuss the novel portions of our code related to the Sweep Plane Data Structure, a data structure we developed to minimize the use of memory during sweeps whilst supporting cyclic dependencies. This is followed by sample results and a scaling study to present our results on meshes and partitions without cyclic dependencies, which we performed out to ~131k processes. We also include a smaller scaling study with global cyclic dependencies to demonstrate the code's ability to handle volumetric partitions, a concept that is more advanced than what is currently seen in literature.
Lastly we discuss our plans for future work and present a conclusion.

\newpage
\section{Brief Background on the Transport Equation and Solution Techniques }
We now present some mathematical background for the implementation of discrete ordinates. For the discussions in this article we will be considering the steady state, multigroup, linear Boltzmann transport equation given by

\begin{equation} \label{eq:baseNTEMG_moments}
\begin{aligned}
\biggr(\Omegabf \cdot \mathbf{\mathbf{\nabla}} +\Sigma_{tg} (\position)\biggr) \psi_{g}(\position, \Omegabf)
&= \sum_{g'=0}^{G-1}
\biggr( 
\sum_{\ell=0}^{L} \sum_{m=-\ell}^\ell
\frac{2\ell+1}{4\pi}  Y_{\ell m}(\Omegabf) \Sigma_{s\ell,g'{\to}g}(\position)
\phi_{\ell m, g'}(\position)
\biggr)
\ + \ q_{g}(\position, \Omegabf), \quad \quad \quad \position\in\mathcal{D}.
\end{aligned}
\end{equation}
\noindent
Here,
\begin{equation*}
\begin{aligned}
\mathcal{D} &=
\text{ the spatial domain of interest with }\partial\mathcal{D}\text{ as boundary}\\
g &= 
\text{is the group index in the list of total groups $G$}\\
L &= 
\text{scattering order or truncation level of the anistropic angular expansion} \\
\psi_g &= 
\text{angular flux for group $g$ in the direction of $\Omegabf$ at $\position$}\\
\phi_{\ell m,g} &=
\text{flux moments for group $g$ at $\position$} \\
\Sigma_{tg} &=
\text{total interaction cross-section for group $g$ at $\position$}\\
 \Sigma_{s\ell,g'\to g} &=
 \text{moment $\ell$ of the scattering cross-section for scattering from group $g'$ to $g$ at $\position$}\\
Y_{\ell m} &=
\text{tesseral spherical harmonics for angle $\Omegabf$}\\
q_g &= 
\text{spatial source for group $g$ in the direction of $\Omegabf$ at $\position$}
\end{aligned}
\end{equation*}
\noindent
The discrete ordinates method involves using a quadrature-method to integrate the product of the angular flux and spherical harmonic functions. A numerical angular quadrature method employs $N$ quadrature points as

\begin{equation} \label{eq:quadrule}
\begin{aligned}
\phi_{\ell m} &= \int_{0}^{2\pi} \int_0^\pi \psi(\varphi,\theta)Y_{\ell m}(\theta,\varphi)\sin\theta d\theta d\varphi \\
&\simeq \sum_{n=0}^{N-1} w_n \psi((\theta,\varphi)_n) Y_{\ell m}((\theta,\varphi)_n)
\end{aligned}
\end{equation}
\newline
where $(\theta,\varphi)_n = \Omegabf_n$ is a discrete angle pair (or ``ordinate") and $w_n$ is the weight associated with this ordinate. The weight is normalized such that

\begin{equation}
\sum_{n=0}^{N-1} w_n = 4\pi
\end{equation}
\newline
\noindent
There are several options for angular quadratures. The classical Level Symmetric quadratures with 90-degree rotation invariance \cite{LewisMiller}, Product quadratures that combine Gauss-Chebyshev quadratures along both the equatorial plane and polar axis (Double Gauss-Chebyshev, DGC), or Gauss-Chebyshev along the equatorial plane combined with Gauss-Legendre (GLC) along the polar axis, and more.

Product quadratures have a distinct advantage in axially extruded meshes which we discuss later. Other quadrature choices include the Quadruple Range (QR) quadrature sets by Abu-Shumays \cite{Abu1}\cite{Abu2}\cite{Abu3}, but although these quadratures are exact for certain spherical harmonics, they only extend to 9 polar and 9 azimuthal points per octant because of the difficulty to determine them analytically at more angles. Jarrel and Adams \cite{LDFE} and later Lau and Adams \cite{LDFE2} introduced Linear Discontinuous Finite Element (LDFE) quadratures which can be used to optimize along certain directions (i.e., refine certain elements on the unit sphere).
\newline
\newline
For the sections that follow we assume we have a set of angles $\Omegabf_n = (\theta,\varphi)_n$ and weights $w_n$ that apply to a given quadrature. We then apply Eq. \eqref{eq:baseNTEMG_moments} to solve each of the abscissae (discrete angles) of the quadrature in the form

\begin{equation} \label{eq:baseNTEMG_moments_discrete}
\begin{aligned}
\biggr(\Omegabf_n \cdot \mathbf{\nabla} +\Sigma_{tg} (\position)\biggr) \psi_{gn} (\position)
&= \sum_{g'=0}^{G-1}
\biggr( 
\sum_{\ell=0}^{L} \sum_{m=-\ell}^\ell
\frac{2\ell+1}{4\pi}  Y_{\ell m}(\Omegabf_n) \Sigma_{s\ell,g'{\to}g}(\position)  
\phi_{\ell m,g'} (\position)
\biggr)
\ + \ q_{gn}(\position)
\end{aligned}
\end{equation}
\newline
where $\psi_{gn} = \psi_g(\Omegabf_n)$ and $q_{gn} = q_g(\Omegabf_n)$. In an iterative sense, the objective is to solve Eq.~\eqref{eq:baseNTEMG_moments_discrete} for each angle $\psi_{gn}$, at iteration $(k+1)$, using flux moments lagged from the previous iteration, $\phi_{g, \ell m}^{(k)}$. The next-iteration flux moments $\phi_{g, \ell m}^{(k+1)}$are updated during the transport sweep, direction by direction and cell by cell. In this manner, the prohibitively memory-expensive angular fluxes need not be stored. This is the basis of Source Iteration applied to transport sweeps.
\newline 
\newline
The transport equation can be cast into operator form as

\begin{equation}
L \psi = MS\phi + Q
\end{equation}
\newline
where $L$ is a block diagonal operator comprising a number of group-wise transport operators $L_g$. $M$ is a moment-to-discrete operator defined from $\frac{2\ell+1}{4\pi}Y_{\ell m}(\Omegabf_n)$ and the scattering operator, $S$, is a diagonal matrix containing the Legendre expansion of the scattering cross sections. This form allows us to easily define a number of iterative algorithms the most simplest of which is the \textbf{Richardson algorithm} (otherwise known as Source Iteration \cite{LewisMiller})

\begin{equation} \label{eq:ClassicRichardson}
\begin{aligned}
\psi^{(k+1)} &= L^{-1} MSD \psi^{(k)} + L^{-1}Q \\
\therefore 
\phi^{(k+1)} &= DL^{-1} MS \phi^{(k)} + DL^{-1}Q \\
\end{aligned}
\end{equation}
\newline
Here the operator $D$ is defined from the weights and spherical harmonics used in the quadrature (Eq. \eqref{eq:quadrule}). Without discussing algorithms that can accelerate the development of the scattering source it will be appropriate to discuss the application of the \textbf{GMRES algorithm}. The source iteration algorithm is simply a fixed point process that is solving a linear system. That linear system matrix is simply obtained by removing the iteration indices and formulating the sweep-preconditioned transport equation as 

\begin{equation} \label{eq:GMRES}
\begin{aligned}
(I - DL^{-1}MS)\phi &=DL^{-1}Q\\
\end{aligned}
\end{equation}
\newline
The GMRES algorithm is a Krylov subspace method introduced in 1986 by Saad \cite{GMRESSaad}; it has been applied to the linear Boltzmann equation since 1998 \cite{GMRES0}\cite{GMRES1}. The performance of the method has been investigated and proven to be very efficient by Patton and Holloway \cite{GMRES2}. Many other subsequent investigations, such as that performed by Adams and Larsen \cite{Adamslarsen}, have demonstrated the efficacy of the algorithm with different preconditioning, the details of which we will not depict here.

In Krylov subspace methods, only the action of the matrix on a (Krylov) vector, i.e., $Av$, is necessary to build the Krylov subspace. The growth of this Krylov subspace can be minimized by using restarted GMRES \cite{GMRESSaad}.

\section{Spatial Discretization using Discontinuous Galerkin Finite Elements on Arbitrary Polyhedral Grids}
The nature of transport type meshes and the need to sometimes define very sophisticated meshes necessitate the implementation of polygons and polyhedrons as the dominant mesh types in Chi-Tech. Note, as we will see later, that polyhedral meshes can also results from a specific type of mesh partitioning to keep sub-domains concave, i.e., without re-entrant faces. The application of the Finite Element Method on these types of cells was inspired by the dissertation of Bailey \cite{PolyFEM} and her more general article on piecewise linear shape functions \cite{Poly2FEM}. In this method the shape functions are constructed by decomposing the cell into simpler constituents, i.e., triangles for polygons and tetrahedrons for polyhedrons.

\begin{figure}[H]
	\centering
	\includegraphics[width=1.0\linewidth]{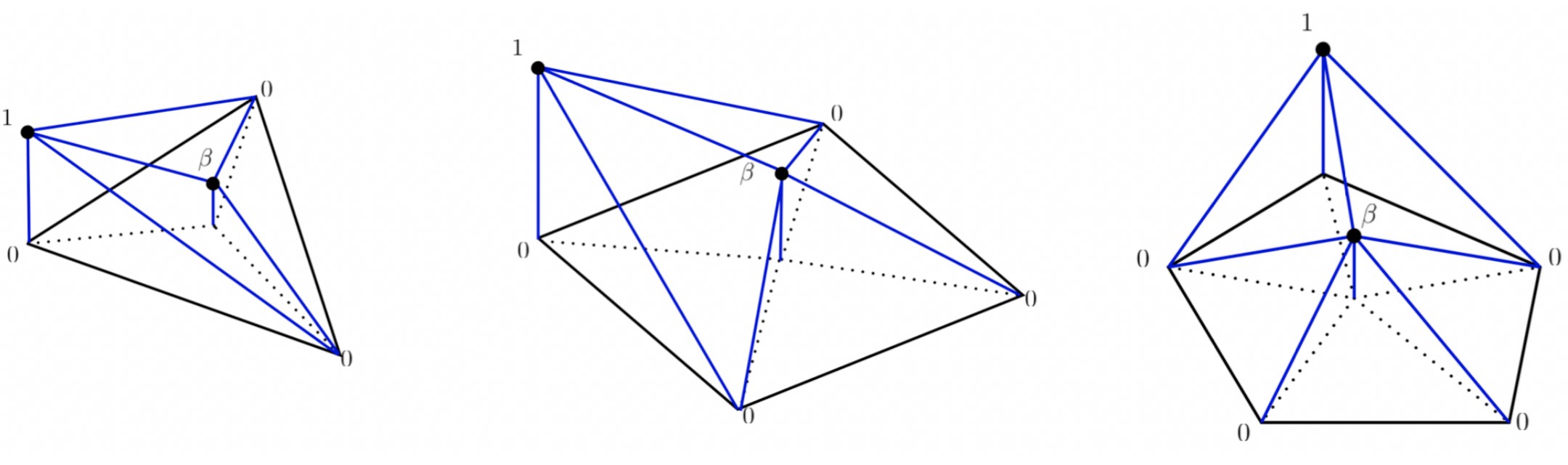}
	\caption{Example shape function application on a triangle, quadrilateral and a general polygon showing the piecewise nature.}
	\label{fig:shapefunctionspolygon}
\end{figure}

For polygons, the triangles are defined by the sides of the polygon and the polygon centroid. As can be seen from Figure \ref{fig:shapefunctionspolygon} one can deconstruct all convex polygonal cells (and most concave polygonal cells) in this fashion. The shape functions are then defined as

\begin{equation}
P_i(x,y) = N_i(x,y) + \beta_i N_c(x,y)
\end{equation}
\newline
where the functions $N_i$ and $N_c$ are the standard linear functions on triangles. The subscript $i$ and $c$ refer to the vertices $i$ and centroid of the polygon, respectively. The $\beta_i$ value is a weighting constant defined such that

\beqn 
\begin{bmatrix}
	x \\ y
\end{bmatrix}_{c}
= \sum_{i=0}^{N_{v}} \beta_i
\begin{bmatrix}
	x \\ y
\end{bmatrix}_{v}.
\eeqn
\newline
Naturally it follows that $\beta_i = \frac{1}{N_v}$ where $N_v$ is the amount of vertices of the polygon. $[x\ y]_{v}$ is the vertex $v$'s coordinate.

\begin{figure}[H]
	\centering
	\includegraphics[width=1.0\linewidth]{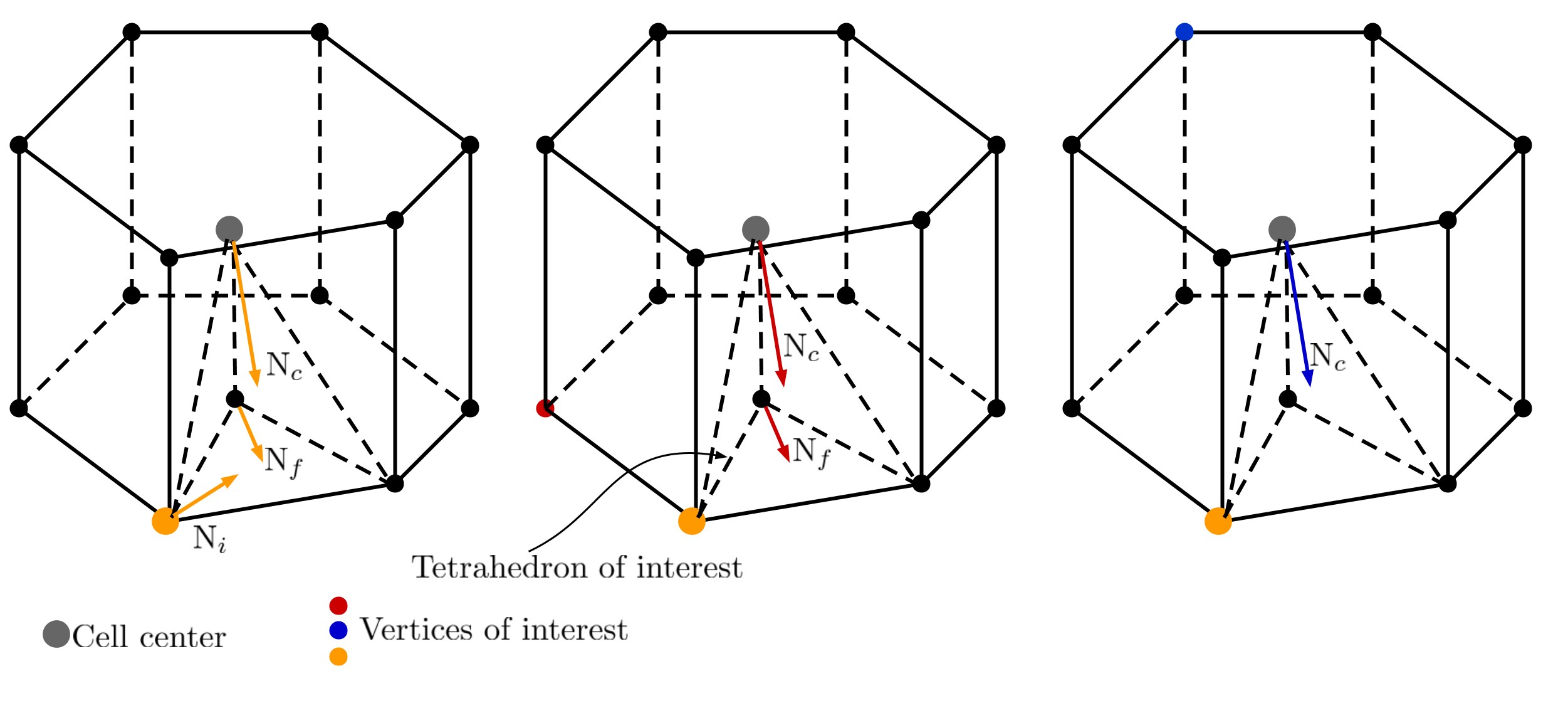}
	\caption{Composition of the piecewise linear shape functions on polyhedron from a reference tetrahedron.}
	\label{fig:threedtetrahedralshape}
\end{figure}

\noindent 
The piecewise linear shape functions on a polyhedron are derived similarly. First we take a face, which itself can be a polygon, then define a tetrahedron as a combination of a side of a face (edge of the face), the face centroid and then the cell centroid (see Figure \ref{fig:threedtetrahedralshape}).

\noindent
The shape functions are then defined as

\beqn
P_i^{tet}(x,y,z) = 
\begin{cases}
	\alpha_c N_c (x,y,z)    \quad \quad &\text{ no matter which tetrahedron} \\
	+\beta_f N_f(x,y,z)  \quad \quad &\text{ if vertex }i \text{ is part of the face} \\
	+N_i (x,y,z) \quad \quad &\text{ if vertex }i \text{ is part of the face-side pair} \\
\end{cases}
\eeqn
\newline
Here the functions $N(x,y,z)$ are the standard linear shape functions defined on a tetrahedron. $\beta_s$ and $\alpha_c$  is the weight that gives the face midpoint, $\bar{r}_{fc}$, and cell mid-point, $\bar{r}_{cc}$, respectively from the sum of the vertices that constitute them. i.e.,

\beqn 
\bar{r}_{fc} =
\begin{bmatrix}
	x \\ y \\ z
\end{bmatrix}_{fc}
= \sum_{v=0}^{N_{vf}} \beta_f
\begin{bmatrix}
	x \\ y \\z
\end{bmatrix}_{v}.
\eeqn

\beqn 
\bar{r}_{cc} =
\begin{bmatrix}
	x \\ y \\ z
\end{bmatrix}_{cc}
= \sum_{v=0}^{N_{vc}} \alpha_c
\begin{bmatrix}
	x \\ y \\z
\end{bmatrix}_{v}.
\eeqn
\newline
where $N_{vf}$ is the number of vertices for the given face and $N_{vc}$ is the number of vertices for the entire cell. Naturally it follows that $\beta_f = \frac{1}{N_{vf}}$ and $\alpha_c = \frac{1}{N_{vc}}$.
\newline
\newline
The weak form of the linear Boltzmann equation can then be derived by multiplying Eq. \eqref{eq:baseNTEMG_moments_discrete} by weight function $\tau_i$ and integrating over a cell volume. Dropping any notations on angle and group, and by defining a generic total spatial source $q$ (external source + scattering contributions) we can derive the weak form of the Boltzmann equation by also expanding the angular flux and spatial source into basis functions $b_j$. We use the same shape functions for the basis functions as the weight function (hence the Galerkin method) and arrive at 

\beqn \label{eq:weakform}
\sum_{j=0}^{N_{dof}-1}
\biggr[
\int_{\partial V} (\Omegabf \cdot \mathbf{n}) \tau_i (\tilde{\psi}_j - \psi_j.b_j).dS  
+\int_V \psi_j \tau_i \Omegabf \cdot  \mathbf{\nabla} b_j .dV
+ \int_V \psi \sigma_t \tau_i b_j.dV
\biggr]= \sum_{j=0}^{N_{dof}-1} \int_V\tau_i q_j b_j.dV.
\eeqn  
\newline
where $\tilde{\psi}$ is upwinded with respect to the dot-product of the angle $\Omegabf$ and the outward pointing surface normal on cell faces (note the entire surface integral term cancels on outgoing faces).

For known upstream fluxes the resulting system forms a square dense matrix of size $N_{dof}{\times}N_{dof}$, the number of degrees of freedom per cell, and since the system is ordinarily small it can be  solved on a cell-by-cell basis using standard Gaussian Elimination.

\section{Parallel Transport Sweeps}
The formulation defined above can be solved using the general notion of a ``sweep" where we start with cells that have all their upstream dependencies met (mainly the boundary cells) and sweep through all the cells, passing downstream angular fluxes. The order in which the cell-by-cell systems are constructed and solved is determined by a \textbf{Sweep Ordering} (SO). In general terms a SO is developed by first constructing a Directed Acyclic Graph (DAG) for a given direction and then applying a topological sorting of this graph. There is a large body of knowledge regarding DAGs from which we specifically used an optimal topological sorting algorithm \cite{TopologicalSorting}. The topological sorting ensures that each cell will be traversed only after its upstream dependencies are met. This is of course subject to the requirement that the dependencies are acyclic (i.e., one-way dependencies). Concave cells, twisted cell faces, parallel partitioning of meshes, and opposing reflecting boundary conditions can create cyclic dependencies for which we will provide a solution for in later sections.
\newline
\newline
For a collection of local cells, the first 6 stages of cells ``swept" are shown in Figure \ref{fig:shapefunctionspolygon1} below. In this figure the stages refer to the number of upstream stages that need to be performed before the cell of the given stage can be solved. The stages are of no algorithmic significance and is used only for this visualization.

\subsection{The Sweep Chunk}
Parallelization of transport sweeps often require the partitioning of the mesh in order to reduce the time of the computation as well as to gain access to distributed memory. In this regime the \textbf{Sweep Chunk} is defined as the unit of computation dealing with cells local to a given processor's partition. In this context we define a simple sweep chunk by the algorithm depicted below:

\newpage
\begin{algorithm}[H]
\For{topologically sorted cell}{
	\For{angles in the sweep ordering, n}{
		//Assemble gradient portion of matrix\;
		\For{i}{
			\For{j}{
				\For{group $g$}{
					$MATRIX A_{i,j}$ $= \int_V \tau_i \Omega \cdot  \mathbf{\nabla} b_j .dV$\;
				}
				
			}
		}
	
		//Reset RHS\;
		\For{i}{
			\For{group $g$}{
				$RHS_{g,i} = 0.0$
			}
		}
	
		//Surface integrals\;
		\For{faces on cell}{
			\If{incident face}{
				\For{FACEDOF i}{
					map i $\leftarrow$ FACEDOF i\;
					\For{FACEDOF j}{
						map j $\leftarrow$ FACEDOF j\;
						$N_{i,j} = \int_{\partial V} (\Omega \cdot \hat{n}) \tau_i b_j.dS$\;
						$MATRIX A_{i,j}$ $+= -N_{i,j} $\;
						\For{group $g$}{
							Obtain $\tilde{\psi}_{gj}$\;
							$RHS_{g,i}$     $\ \ += -\tilde{\psi}_{gj}   N_{i,j}$\;
						}
					}
				}
			}	
	    }
    
    	\For{group $g$}{
    		//Compute source moments\;
    		\For{DOF i}{
    			$q_{src}$ = 0.0\;
    			\For{moment $m$}{
    				//Moment-to-discrete operator
    				$q_{src} += M_{n, m}\times q_{mom}$\;
    			}
    			$q_i = q_{src}$
    		}
    		//Assemble mass matrix portion into temporary matrix and source moments\;
	    	\For{DOF i}{
	    		temp = 0.0\;
	    		\For{DOF j}{
	    			$\mathcal{M}_{i,j} = \int_V \tau_i b_j.dV$\;
	    			$MATRIX A_{i,j}^{temp} = A_{i,j} + \sigma_{tg} \mathcal{M}_{i,j} $\;
	    			temp $+= q_j \mathcal{M}_{i,j} V$
	    		}
    			$RHS_{g,i}$ += temp\;
	    	}
    	
    		Gauss-Elimination $\to$ $A^{temp} x_g = b_{g}$
   		}
    	
    	//Accumulate flux moments\;
    	\For{moment $m$}{
    		\For{DOF i}{
    			\For{group g}{
    				//Discrete-to-moment operator\;
    				$\phi_{g,m,i} \ += D_{m,n} x_{g,i}$
    			}
    		}
    	}	
    
    	Store outgoing fluxes
    }	
}
\caption{Sweep Chunk}
\end{algorithm}

\newpage
\noindent
The above algorithm has many details that are worth discussing. Firstly, each sweep operation is prefaced by a major routine that computes $q_{mom}$ on each degree of freedom on each cell. Even though it is a major step, it has significantly fewer unknowns because it is angle independent and therefore still dwarfed by the sweep chunks. In this routine, the ``source-update" routine, the transfer matrices are applied to obtain scattering moments, the fixed source is applied and an isotropic production source is applied if present (i.e., fission), and is collected into the source moments, $q_{mom}$. 

Within the sweep chunk, the items of interest are firstly the matrix $A$ and the matrix $A^{temp}$. The $A$ matrix is defined for each cell-direction and its group independent entries, i.e., gradient and surface integral terms, are precomputed. For each group $g$ we then only recompute $A^{temp}$, to which $A$ is added before the Gaussian-Elimination step for each group. We recompute the right hand side vectors on a group-by-group basis every time. The temporary matrix saves a lot of computation. Also note that $\mathcal{M}$, representing the mass-matrix element, should not be confused with the moment-to-discrete operator $M$. Note also that we define multiple right-hand side vectors, one for each group, which allows us some flexibility in looping over groups.
\newline
\newline
In terms of computational cost the sweep chunk is the major cost item and represent the ``Task" that each process must perform for a given collection of angles. The notion of ``collection of angles" is rather vague and will be explained in later sections. The other important item is the ``Obtain $\tilde{\psi}$" step in the face integrals portion and the final step where we store outgoing angular fluxes. Both these items are the topic of one of our next sections since they require a very important data structure, however we can add the remarks below.
\newline 
\newline
Incident fluxes can be obtained from the following sources:
\begin{itemize}
	\item Incident flux boundaries,
	\item Cells executed in the local sweep ordering with ranks less than the current cell,
	\item Upstream partitions, executed in the global sweep ordering before the current partition's execution,
	\item Reflecting boundary conditions, and finally
	\item Cyclically dependent delayed upstream fluxes (either from local cells, neighboring partitions or opposing reflecting boundary conditions)
\end{itemize}
\noindent
Outgoing fluxes need to store their data conducive to the incident flux mappings, descriptions of which are included in the sections that follow.

\begin{figure}[H]
	\centering
	\includegraphics[width=0.8\linewidth]{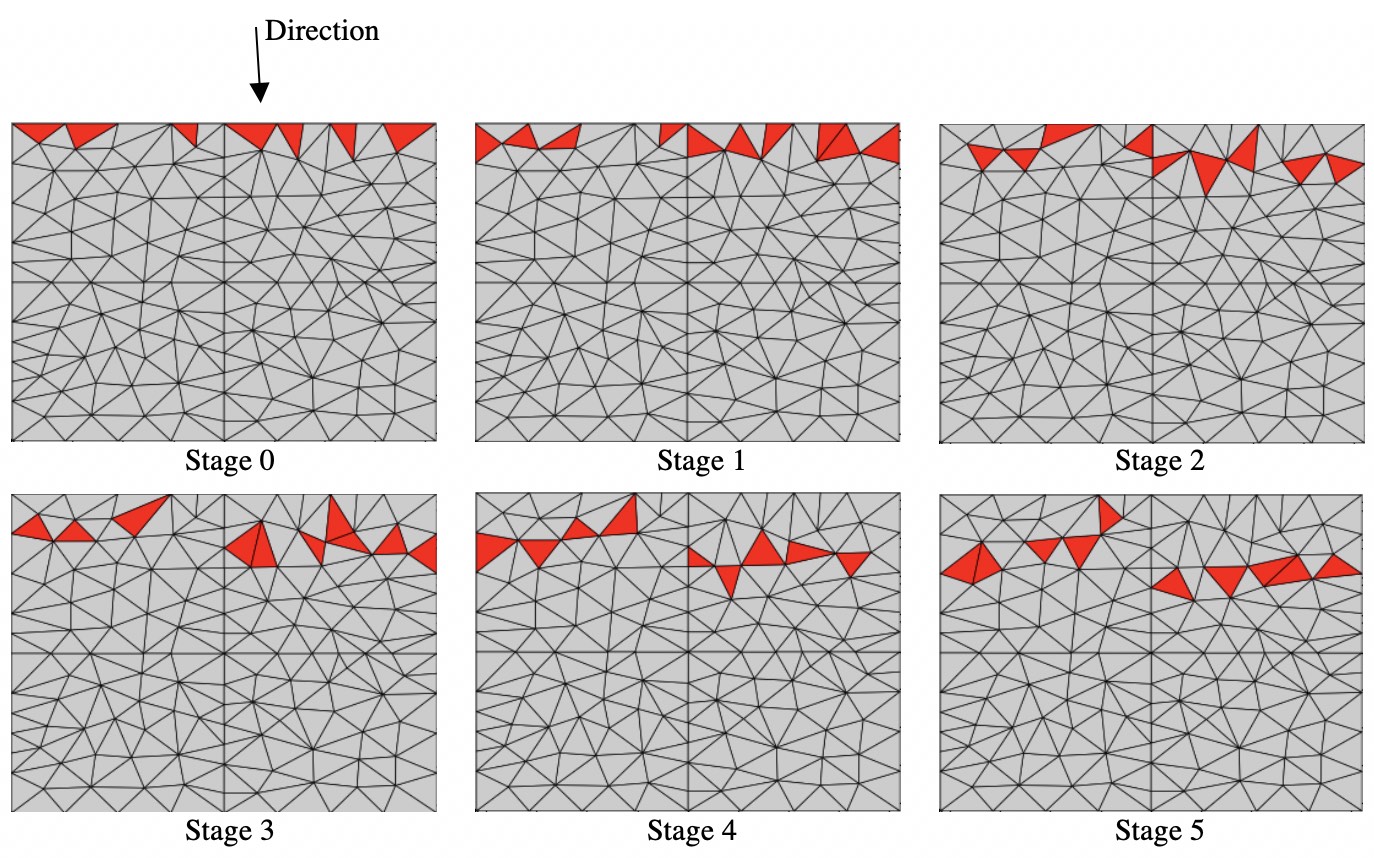}
	\caption{The first 6 stages of a sweep operating on an unstructured 2D triangular mesh.}
	\label{fig:shapefunctionspolygon1}
\end{figure}

\newpage 
\subsection{Sweep timing}
Apart from parallel efficiency we measure the timing of sweep operations as the average time taken to sweep across the whole domain (excluding the source update routine), $T_{avg}$, divided by the total number of angular unknowns. That is, for a given sweep, $N_{unknown} = N_{DOF,total} \times N_{angles} \times N_{groups}$ is the total amount of unknowns ($N_{groups}$ is not necessarily the total number of groups, $G$, in the simulation as discussed later), and the sweep timing (called the ``grind time") is given by

\begin{equation}
T_{grind} = \frac{T_{avg}}{N_{unknown}}
\end{equation}
\newline
For a piecewise linear spatial discretization we observe grind times of less than 100$ ns$ per angular unknown with optimal aggregation parameters. This is also dependent on CPU architecture. More performance detail will be given on this topic later.

\subsection{Workloads and CPU caching effects}
Within the sweep chunk we have the cell-by-cell outer loop followed by the angle-by-angle loop and then the internal structure containing the group-by-group loops. It is not intuitive to think that the group-by-group loops in the internal structure is beneficial, but there is a caching effect on CPU workload that has to be considered. The primary effect here is that the data associated with cells gets pipelined and cached in such a manner that the CPU can manipulate data structures more efficiently when the data is more contiguous. This is indeed the case when considering the mapping of cell degree of freedom information where the blocks of data is closer together in memory once a cell is executed. This is why the loop over cells is the outer most loop. This argument holds true for the flux moments and for each group of the flux moments.

We have seen tremendous improvements in Chi-Tech's sweep chunk performance by arranging our degree of freedom data structure to be contiguous in groups and then executing as many groups in a sweep chunk as possible. We call the process of adding a collection of groups to sweep chunks \textbf{group aggregation} and the serial performance of the sweep chunk indicates that grind times reduce with increasing amount of groups aggregated.

Angle aggregation is discussed later but also has an effect on serial sweep chunk performance. 
Additionally, aggregating groups, angles and cellsets (cellsets are a PDT-only feature) has the effect of increasing the ratio of task-time to communication time which inherently increases parallel performance but can be offset by other effects such as pipe-fill and drain discussed in later sections.

\subsection{Groupsets and groupset subsets}
With the tremendous amount of groups sometimes involved in transport calculations, it is often advantageous to converge sets of groups first before converging others. In both PDT and Chi-Tech there is the concept of groupsets. In a typical 172 group WIMS multigroup structure \cite{WIMS}, the total amount of groups can require too much memory to execute sweeps on. In these cases the strategy is often to lump groups into sets that are converged one at a time. This means we apply Source Iteration or GMRES only on the groupset until converged. Even though this can be advantageous, it can sometimes pose problems with regards to upscattering calculations. In other words if we lump groups into groupsets of 20 groups each then we can easily converge groupsets starting from sets containing high energy groups to groupsets containing lower energy groups (in some ways one can think of this operation as a block Gauss-Seidel scheme). The more common strategy is to split the down-scattering dominated groups into a few groupsets and then one large groupset for the thermal region exhibiting up- and down-scattering. Since the memory requirement is reduced for the smaller high energy groupsets we can increase the amount of angles used in the quadrature, an aspect that assists with the highly anisotropic flux nature of these groups. Conversely, less memory is available in the thermal energy groupset which requires fewer angles in the quadrature, however, these energies are more likely to exhibit a more isotropic flux and therefore not as sensitive to the number of angles used.
\newline
\newline
Within a sweep chunk we only execute the groups within a given groupset. Since the higher energy groups are mainly dominated by down-scattering we achieve very fast convergence and often without the need for computationally expensive Diffusion Synthetic Acceleration (DSA) as preconditioning. DSA schemes exist to accelerate the within-group scattering sources \cite{WGDSA2} \cite{WGDSA},  and the group-to-group thermal upscattering scattering sources as developed by Morel \cite{TwoGrid} (i.e., Two grid acceleration). Specifically we used Turcksin and Ragusa's Modified Interior Penalty method \cite{MIP} for the spatial discretization of both schemes and solve the systems using PETSc's \cite{PETSc} Conjugate Gradient iterative solver with HYPRE's BoomerAMG \cite{HYPRE} preconditioner. With the splitting among groupsets we can selectively apply DSA to low energy groupsets to reduce the simulation's overall memory footprint.

Given a sweep to be performed on a specific groupset, another strategy to increase the pipe-length (necessary to reduce pipe-fill and drain) is to create groupset subsets. This involves no change in flux data structures (discussed later) for a given sweep setup and rather just increases the amount of angle-sets. 

\subsection{Domain partitioning and parallel sweeps}
\begin{figure}[H]
	\centering
	\includegraphics[width=0.4\linewidth]{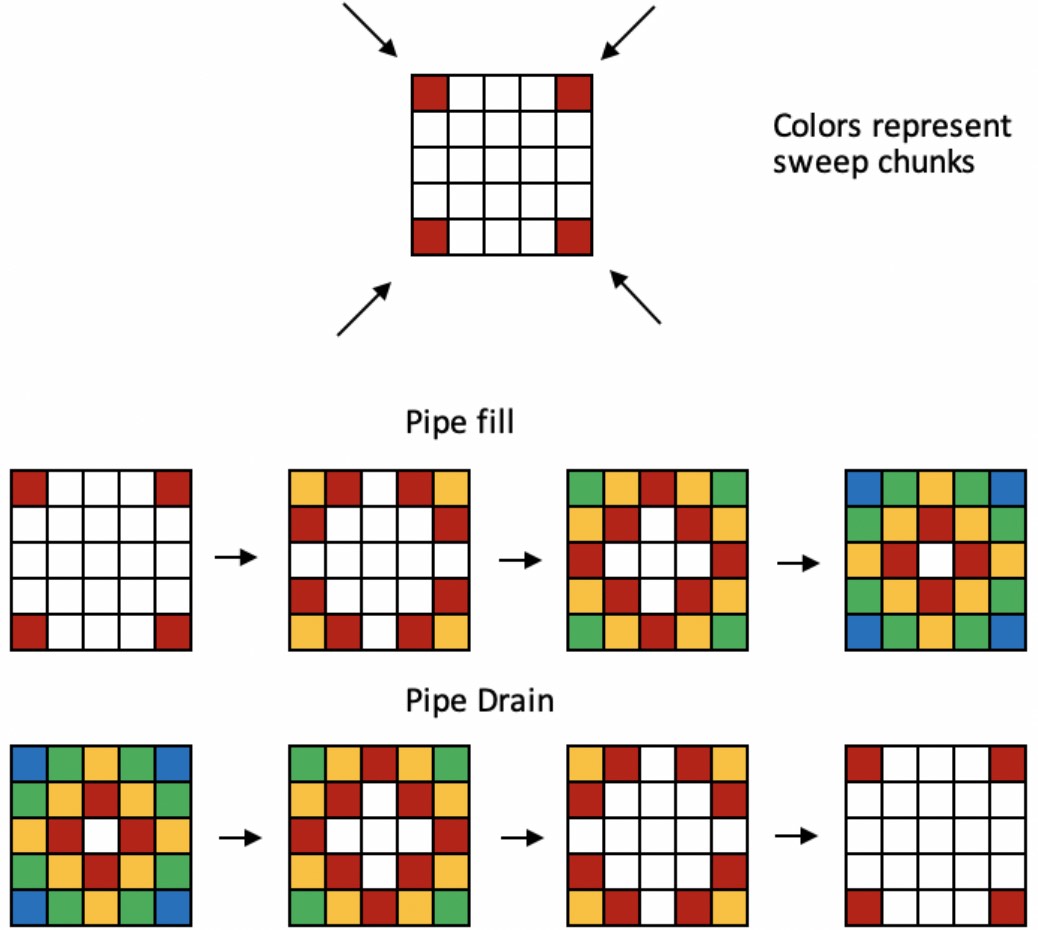}
	\caption{Pipe fill and drain phenomenon associated with a $P_x$-$P_y$ partition layout.}
	\label{fig:pipefilldrain}
\end{figure}

Because of the tremendous amount of unknowns in a transport simulation (i.e., product of spatial degrees of freedom, groups, and moments) it is essential to properly decompose the spatial domain during parallel simulations. The paper by Bailey et al. \cite{ValidationSweeps} described the problems associated with sweeps and these parallel decompositions. The primary concern is that processors experience idle time as the sweep operations move from partitions with satisfied dependencies to partitions waiting on dependent partitions. This phenomenon is sometimes called ``pipe fill" and has a similar effect during ``pipe drain" as shown in Figure \ref{fig:pipefilldrain}. 

The first algorithmic item to implement is that sweeps along angles should be able to start from any corner of the domain. This is possible with little trouble and should be done to keep pipe-fill and drain to a minimum. The question then arises of what to do when all the angles arrive at a given partition (i.e., the center partitions) and how to apply tie-breakers. The recent paper by Adams on provably optimal parallel sweeps \cite{ProvablyOptimal} depicts the Depth-Of-Graph (DOG) sweep scheduling algorithm where sweep tasks are prioritized, on a partition, by means of their depth in the graph associated with a given angle. The paper also details how to handle tie breakers. This sweep scheduling algorithm has been adopted in Chi-Tech and determined to be superior to a simple First-In-First-Out (FIFO) scheduling. 

The Adams performance model for semi-structured meshes provides an equation for the theoretical parallel efficiency, $\epsilon_{opt}$, which we have modified slightly to match our definitions as 

\begin{equation} \label{E_opt}
\epsilon_{opt} = \biggr[ 
\biggr(
1 + \frac{P_x + \delta_x +P_y + \delta_y + P_z + \delta_z -6}{\omega_{ang}}
\biggr)
\biggr(
1 + \frac{T_{comm}}{T_{task}}
\biggr)
\biggr]^{-1}
\end{equation}
\newline
where $P_x$, $P_y$ and $P_z$ are the brick-type partitioning layout of the domain, the $\delta$ factors are 1 if the corresponding $P$ factor is odd and 0 if even. The aggregation factor $\omega_{ang}$ is the total amount of anglesets (defined in the next section) across all the octants. $T_{comm}$ is the time taken to communicate angular fluxes between partitions and $T_{task}$ is time taken per execution of a sweep chunk.

\noindent 
Mesh partitioning using the KBA method poses some challenges with regards to non-orthogonal meshes. This is because orthogonal meshes can be partitioned with x-, y- or z-slices without introducing cyclic dependencies between partitions (described in later sections on global cyclic dependencies). In order to maintain a cycles free partitioning on unstructured meshes it is required to ``cut" the mesh along the cut planes used for partitioning. This invariably introduces more cells to the problem.
 The method of cutting has been used for many scenarios by PDT and has proven to be a successful method to partition a simulation domain. Since PDT specializes on extruded meshes the code only applies the cutting method to x- and y-cuts.
 
 Volumetric partitioning involves the process of collecting cells into contiguous groups without modifying the original cells. These groups can be created by various means. One such method is to use the x, y, and z-planes that were used to define a KBA-style partitioning (resulting in $P_x{\times}P_y{\times}P_z$ brick shaped volumes) but instead of cutting the mesh, we include cells in a partition if their centroids are within the corresponding brick shaped box. An example of this is shown later. Algorithms such as those implemented in ParMETIS \cite{ParMETIS} offer additional flexibility by optimizing the partition for certain types of iterative methods.  Because volumetric partitioning does not modify cells, the resulting partitions from an unstructured mesh are likely to have concave faces, an aspect that we will later show to cause cyclic dependencies.

\subsection{Anglesets and polar angle aggregation}
An angleset is a collection paradigm for aggregated quantities. All angles which share a sweep ordering can be aggregated into an angleset. For orthogonal grids this is usually possible for all the angles in an octant. Additionally group aggregation is also included in this paradigm with one small modification, i.e., only the groups in the groupset-subset are included in an angle set. Both these aggregation types afford us ways to reduce the memory footprint of data structures whilst simultaneously affecting performance.
Extruded meshes offer many advantages in transport simulations combined with product quadratures that have the same polar angles for each azimuthal angle. This is because the top and bottom hemispheres of the quadrature, respectively, will have the same sweep ordering for all the polar angles in that hemisphere and therefore we have the option to aggregate the polar angles into angle sets. In this work we focused mainly on extending our methods to completely unstructured meshes and therefore applied only \textbf{single angle aggregation}. However, the performance related aspects need to be examined carefully when implementing such methods. In particular it has 3 main effects:
\begin{itemize}
	\item It decreases the amount of tasks (i.e., sweep chunks) to be performed per partition. This decreases the pipe length and therefore can reduce parallel efficiency, however,
	\item It increases the time per task and therefore reduces ratio of communication time to task execution time. This increases parallel performance. Lastly,
	\item It has a tremendously beneficial caching behavior resulting in significantly increased serial performance which might be more significant than the overall parallel efficiency.
\end{itemize}

We reiterate that the general concept of an angleset includes the angles and groups (in the form of groupset subsets) aggregated into the angleset.

\subsection{Local and global cyclic dependencies}

At the time of this writing there are only two effective strategies for dealing with cyclic dependencies. The first and simplest strategy is to remove the cause. If we do not have concave cells (with re-entrant faces), do not apply volumetric partitioning (which induce re-entrant partitions),  and follow a different strategy for opposing reflecting boundary conditions, then clearly the problem would not exist. Concave cells can be avoided by ensuring concave cells are not present during mesh generation. High quality mesh generators normally ensure this. Indeed this strategy could also involve the splitting of curvilinear edges/faces to straighten them which was applied in the paper by Haut et al. \cite{EdgeSplitting}. One could also apply a sophisticated cell splitting scheme as was devised by Liu and Larsen \cite{CurvedGrids} for curved meshcells. 

The second strategy for dealing with cycles is to selectively assign delayed storage to one side of a cyclic dependency. That is, if cell A and cell B are cyclically dependent, allow cell A to pass flux data to cell B but let cell B store its outgoing data (upwind flux for cell A) in a structure that is one iteration behind. Cell A then reads from the ``delayed" angular flux data instead of the concurrent data from cell B. In this fashion one can effectively remove the cycle. This concept is known as ``detecting and breaking cycles". The same corollary also applies for cyclically dependent mesh partitions (global cyclic dependencies), such as those arising from volumetric partitioning, and for opposing reflecting boundary conditions. 

For global cyclic dependencies either the incoming- or outgoing message buffers can be classified into old and new data. We applied this classification to the incoming message buffers in order to ensure that a location has old information available as soon a new sweep is executed. New information is communicated after each sweep. 

Opposing reflecting boundary conditions follow the same paradigm but the difference here is that the dependencies are reflected-angle based and local to a partition. Single (non-opposing) reflecting boundary conditions can be accommodated by the sweep scheduler. Anglesets who are awaiting their reflected angle are delayed until their reflecting angle has written to its boundaries. Therefore, when adding the second opposing reflecting boundary condition, only the second boundary needs special treatment with regards to cyclic dependency. The treatment of single reflecting boundary conditions allows the parallel nature of the simulation to mimic a larger domain. Using the scheduler property, the PDT simulations performed by Adams et al. \cite{ProvablyOptimal} on $\simeq$ 770k cores using a IBM BG/Q computer, was extended to the equivalent of 1.5 million processes. 

This scheduling paradigm is also implemented in Chi-Tech, with the addition of the opposing reflecting boundary conditions (not implemented in PDT).

The graph based algorithms in Chi-Tech compares very closely to the algorithms developed in the paper by Plimpton \cite{RemoveCycles}. The same group also authored a paper on how to find Strongly Connected Components (SCCs), a term widely used to refer to cyclic dependencies. In a LLNL document by Haut et al. \cite{RemoveCycleLLNL} they also remove the cycles from graph edges and delay the information.

A succinctly described formulation of the iterative process was found in a paper by Wang and Ragusa \cite{RemoveCyclesYaqi} from which we will formulate our implementation of the GMRES iterative scheme with reference to Eq. \eqref{eq:GMRES}. We start with the inclusion of all cyclically dependent angular unknowns, $\psi^{c}$, in the original angular flux vector $\psi^* = [\psi \quad \psi^{c}]^T$. We then increase the size of the transport operator and name it $L^*$ after which we have

\begin{equation}
\begin{aligned}
L^*
\begin{bmatrix}
\psi \\
\psi^{c}
\end{bmatrix}
 - 
\begin{bmatrix}
MSD & 0 \\
0 & 0
\end{bmatrix}
\begin{bmatrix}
\psi \\
\psi^{c}
\end{bmatrix}
=
\begin{bmatrix}
Q \\
0
\end{bmatrix}
\end{aligned}
\end{equation}
\newline
We then split our transport operator into its block lower triangular component, $L_L$, which represents the non-cyclically dependent portion of $L^*$, and its block upper triangular component, $L_U$, which represents the cyclically dependent portion.

\begin{equation}
L^* = L_L - L_U
\end{equation}
therefore 
\begin{equation}
\begin{aligned}
L_L \begin{bmatrix} \psi \\ \psi^{c} \end{bmatrix}
- \biggr(
L_U \begin{bmatrix}
M & 0 \\
0 & I
\end{bmatrix}
- \begin{bmatrix}
MS & 0 \\
0 & 0
\end{bmatrix}
\biggr)
\begin{bmatrix} \phi \\ \psi^{c} \end{bmatrix}
=
\begin{bmatrix} Q \\ 0 \end{bmatrix}\\
\end{aligned}
\end{equation}
\newline
We then invert the lower triangular operator by sweeping and formulate a new GMRES iterative scheme as

\begin{equation}
\begin{aligned}
\biggr( I
- \begin{bmatrix}
D & 0 \\
0 & I
\end{bmatrix}
L_L^{-1}\biggr(
L_U \begin{bmatrix}
M & 0 \\
0 & I
\end{bmatrix}
- \begin{bmatrix}
MS & 0 \\
0 & 0
\end{bmatrix}
\biggr)
\biggr)
\begin{bmatrix} \phi \\ \psi^{c} \end{bmatrix}
=
\begin{bmatrix}
D & 0 \\
0 & I
\end{bmatrix}
L_L^{-1}\begin{bmatrix} Q \\ 0 \end{bmatrix}\\
\therefore
A^* \begin{bmatrix} \phi \\ \psi^{c} \end{bmatrix} = b^*
\quad \quad \quad \quad \quad \quad \quad \quad \quad
\quad \quad \quad \quad 
\end{aligned} 
\end{equation}
\newline
The block matrices in this formulation are the equivalent of the result of a sweep chunk, i.e., the matrix-vector multiplication occurs in a more implicit sense than depicted here. For GMRES the right-hand side vector, $b^*$ gets precomputed. Thereafter the matrix action requires a sweep to obtain

\begin{equation}
w = 
\begin{bmatrix}
D & 0 \\
0 & I
\end{bmatrix}
L_L^{-1}\biggr(
L_U \begin{bmatrix}
M & 0 \\
0 & I
\end{bmatrix}
- \begin{bmatrix}
MS & 0 \\
0 & 0
\end{bmatrix}
\biggr)
v
\end{equation}
\newline
which is a standard sweep followed by accumulating the flux moments (not the cyclically dependent angular fluxes). The additional result is that cyclic dependent angular fluxes are also updated and directly stored. The terms inside the parentheses are the scattering sources and cyclic face fluxes based on the Krylov vector $v$ and is available to a sweep chunk. 

\newpage
\section{Sweep Plane Data Structure}
The aspect of Chi-Tech that pushes the state-of-the-art in terms of performance is the Sweep Plane Data Structures (SPDS). Given the angles in a quadrature, defined per groupset, we proceed with constructing the equivalent of local- and global Directed Acyclic Graphs (DAGs). During this phase we identify and break cyclic dependencies in the following fashion.
\newline
\newline
\textbf{Local sweep ordering (cells on a partition):}
\begin{itemize}
	\item We determine cell connectivity. If a cell is found to be strongly connected to another cell, one of the cells (typically the first one of the pair encountered) will have its dependency on the other cell marked as ``delayed".
	\item By virtue of the same loop over local cells we will ultimately also encounter the partition boundaries. We identify partition successors and predecessors during the collection of these events.
	\item We then topologically sort the cells in the DAG and obtain the local sweep ordering.
\end{itemize}
\noindent
\textbf{Global sweep ordering (partition level):}
\begin{itemize}
	\item In the previous step we determined the partition-wise successors and predecessors which in itself is a form of a DAG.
	\item We broadcast these dependencies to all partitions so that each process has the full partition-wise DAG.
	\item We then perform a topological sorting of the DAG to determine the global sweep ordering.
\end{itemize}
\noindent
The global sweep ordering is performed on the same DAG on each process. This could have been problematic if we used this sweep ordering for communication because the topological sorting is not guaranteed to be unique, however, we use this sorting only in the Depth-of-Graph scheduling algorithm to assign task priorities.

\subsection{Flux Data Structure (FLUDS)}
The next step in the algorithm is to develop Flux Data Structures (FLUDS). Each angleset has a FLUDS which allows us to store indexes to the locations where outgoing faces will store data and where incoming faces will read it from. This step is prefaced by a very import and novel concept. The faces of the local cells are sorted into a histogram representing the number of degrees of freedom per face. In a simple mesh, where all the cells are hexahedra, the histogram will be flat with only one category (namely 4 DOFs per face). In a polyhedral mesh where we might have degenerate cells (a few cells with a large amount of DOFs per face), the histogram is categorized as follows:
\begin{itemize}
	\item A list of DOFs-per-face is constructed and sorted from smallest to largest. If there are F faces, there will be F list items. (i.e., 3,3,3,3,3, 4,4,4,4,4,4,4,4, 5,5,5,5,5,5)
	\item The average and maximum DOFs per face is computed.
	\item If the maximum-to-average ratio is $\le 1.2$ (can be customized) then all the faces are grouped into one bin designated to have a total number of face DOFs less than or equal to the maximum DOFs per face and the next step is skipped.
	\item If the maximum-to-average ratio exceeds 1.2 then the following algorithm is applied:
	\begin{itemize}
		\item A running average, $n_{avg}$ is instantiated with initial value set to the smallest amount of DOFs per face. The current face size is also set to the smallest amount of DOFs per face ($f_{size}$).
		\item The sorted face list is traversed (for each face $f$)
		\item If the ratio of the number of DOFs on face $f$ to the running average exceeds 1.1 (can be customized) then a new bin is created. The bin represents the faces with number of DOFs less than or equal to the number of DOFs currently stored in $f_{size}$ and greater than the associated size limit of the previous bin.
		\item $f_{size}$ is reassigned to the value of face $f$
		\item After all faces are traversed, a final bin is added which represents all faces with number of DOFs smaller or equal to the maximum number of DOFs per face and greater than the size associated with the previous bin (if it exists). This last step also covers the case where the maximum to average ratio is less than 1.2.
	\end{itemize}
\end{itemize}
\noindent
With this face histogram developed we can now build a FLUDS which can store faces into reusable ``spots" or slots. We instantiate a number of ``lockboxes", one for each face category. Each lockbox is initially empty but earmarked to store the number of face DOFs associated with its category (times the number of angles and groups). We also instantiate another set of lockboxes, this set is however designated to hold delayed angular fluxes.

The next step in the algorithm here is quite novel. It is split into an $\alpha$-phase and $\beta$-phase. Both phases require a loop over cells and faces in the sweep ordering, however, the first phase cannot complete the entire data structure because it requires information about neighboring cells on neighboring partitions. The second phase is therefore prefaced by communicating information about cells and faces that are on partition interfaces.
\newline
\newline
\textbf{FLUDS $\alpha$-phase}:
During this phase we loop through the local cells according to the local sweep ordering associated with the angleset. We then loop over the faces of the cell and if a face is outgoing we firstly identify the face category and therefore its associated lockbox. Note that a face category could be a delayed lockbox category too, if the cell marked its successor as a cyclic dependency. We identify a location inside of the lockbox. If the lockbox is empty, we add a slot and fill it with a pair of indices: a cell index, identifying which cell wrote to the slot, and a face index. This will give us an index into where this face will store its angular fluxes (contiguous, in the order of the DOFs). We store this index.

If a face is incident we determine if it needs to read from a local cell. If it does we find the lockbox slot where that cell and face pair was designated to store its data. We empty that slot making it available to write to again. Because we are looping through the true sweep ordering we are assured that we will not be overwriting data when a slot is reused since the incident face of a dependent cell will read from this slot before another face will write to it. With this portion completed, each face knows where to write its outgoing fluxes and where to read it from. Additionally, since we are reusing slots, we are saving a tremendous amount of memory because the maximum amount of angular fluxes to store only occurs when the maximum amount of cells are ``in-play".

The last portion of the $\alpha$-phase is to map incident faces to their upstream counterparts. This is fairly simple and required since the incident face will have opposite DOF ordering according to the right hand rule.

A graphical representation of the lockbox concept is shown in Figure \ref{fig:lockbox} below.
\newline
\newline
\textbf{FLUDS $\beta$-phase}: During the $alpha$-phase outgoing faces would have identified where and in which lockboxes it stores its outgoing data. This also holds true for the angular fluxes destined for dependent partitions. During this phase the entire lockbox information is sent to- and received by the dependent locations, including information on the DOFs on each face. This is a simple serialization and deserialization process. Once the dependent location has received this information it can create an appropriate mapping of data on incident faces associated with its predecessor locations. This portion is done as a ``Pseudo" sweep and is very fast.

The last portion of this phase is to receive lockbox and cell information relating to delayed predecessors. This cannot be done in a pseudo sweep as before and is rather done in a broadcast sense.

\begin{figure}[H]
	\centering
	\includegraphics[width=0.7\linewidth]{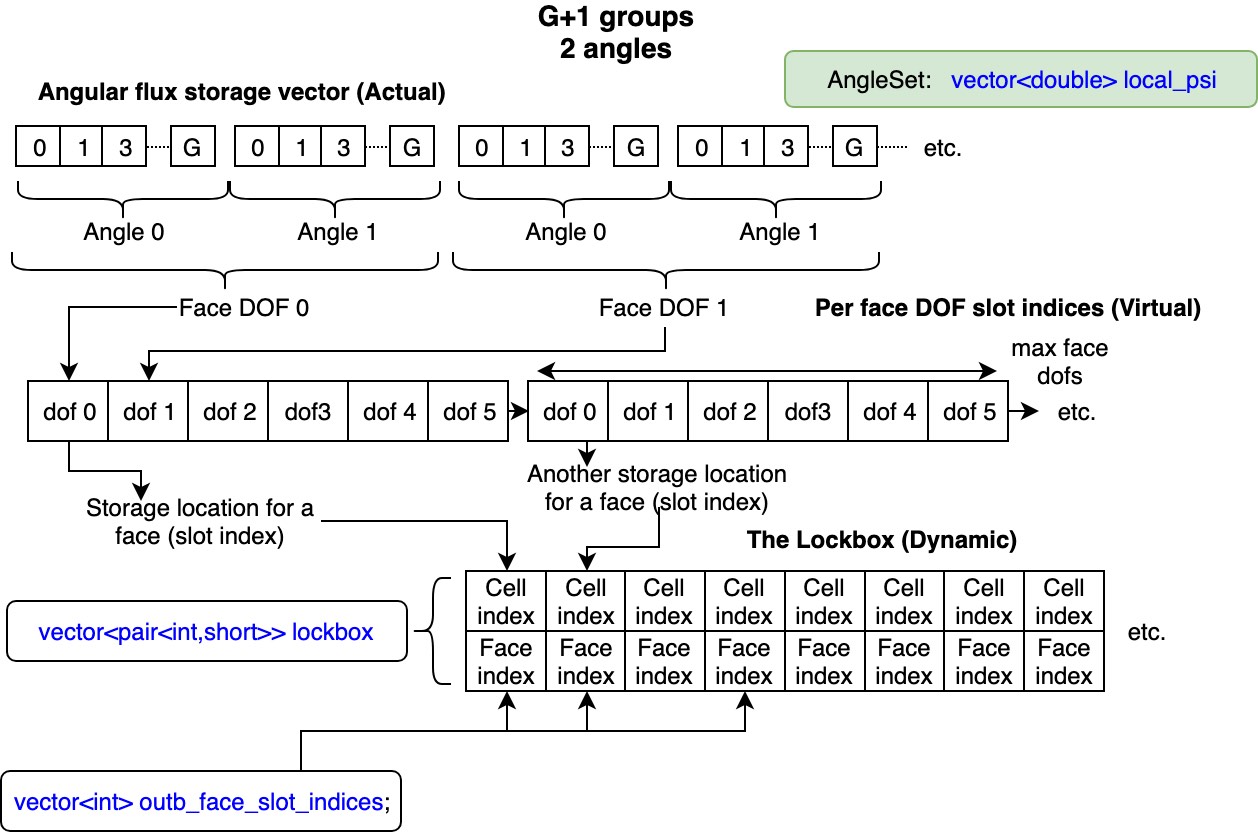}
	\caption{Graphical depiction of the ``lockbox" concept.}
	\label{fig:lockbox}
\end{figure}

\noindent 
At the conclusion of each FLUDS initialization. Each angleset will have storage for the following (in different states of initialization):
\begin{itemize}
	\item Local angular fluxes
	\item Delayed local angular fluxes, old and new
	\item Angular fluxes destined for successor partitions
	\item Angular fluxes destined for delayed successors
	\item Angular fluxes received from predecessor partitions
	\item Delayed Angular fluxes received from predecessor partitions, old and new
\end{itemize}

Opposing reflecting boundary conditions store their angular fluxes (delayed and non-delayed) on their on data structure and will not be explained here.

\subsection{Sweep Buffers}
The sweep buffers are objects operating on anglesets and encapsulate all the MPI related communication. In particular, they are responsible for breaking up communication into packets that fit into the given architecture's communication buffers (sometimes referred to as the ``eager" limit for programs using MPI), and appropriately allocating and de-allocating memory.

During an actual sweep the sweep buffer will size the receive buffers and repeatedly listen for signals to receive data (receive it packet-by-packet as necessary). Once all the dependency communication is completed an angleset will be marked as ready-to-execute. If the sweep scheduler determines that it has sufficient priority, it will signal the sweep buffer to allocate local memory and command the angleset to execute its sweep chunk. Upon completion of the sweep chunk the sweep buffer will clear all the local non-delayed angular fluxes and receive buffers (freeing memory), and it will initiate communication with successor locations. It will continue to monitor the status of communication until it ascertains that successor locations received all the data (or the local storage is no longer necessary) upon which it will clear the send buffers. At this point all of the memory for the current angleset would have been released back to the system. 

The low-overhead of Chi-Tech simulations (using a large amount of processes) can be attributed to the machinery of the sweep buffers. The sweep buffer operates on locally defined communicators and overlaps communication in an efficient manner that avoids unnecessary delays. It operates directly on the data structures without the need to pack data to-and-from graph data structures as is done in PDT-STAPL implementation.

\section{Sample large-scale simulation results}
A visualization of a sample large scale simulation is shown in Figure \ref{fig:samplesim} below. The simulation modeled an experiment used in an uncertainty quantification study by Texas A\&M's Center for Exascale Radiation Transport \cite{CERT}, and contained a large amounts of thermalizing material which included graphite, high density polyethylene (HDPE) and borated HDPE. An Americium-Beryllium neutron source encased in stainless steel was used to supply fast neutrons. The simulations were compared to measurements made using a spherical BF$_3$ detectors which were modeled semi-full detail (without thin stainless steel shell). The spatial domain was discretized with a mesh containing $\simeq$540k cells. Materials were modeled using 116 energy group cross-sections generated by the SERPENT2 code \cite{SERPENT}(scattering order $L=3$). The angular discretization used a Gauss-Legendre-Chebyshev (polar-azimuthal) quadrature with 4,608 angles (96 polar, 48 azimuthal angles) for the fast groupset (52 groups), and 2,688 angles (48 azimuthal, 56 polar) for the thermal groupset (64 groups).

The simulation was performed with 2,204 processes on LLNL's Quartz machine, with a load balance factor of 2.2. Total simulation time was approximately 2 hours and 30 minutes, grind times were $\simeq$298$ns$ per angular unkown for the fast groupset and $\simeq$314$ns$ per angular unknown for the thermal groupset.

For the thermal groupset, within-group Diffusion Synthetic acceleration (WGDSA) was used along with Two Grid Diffusion Synthetic Acceleration \cite{TwoGrid}. The acceleration algorithms used the Discontinuous Finite Element Method along with the Modified Interior Penalty method \cite{RemoveCyclesYaqi}. They were solved using PETSc's Conjugate Gradient method preconditioned with HYPER's BoomerAMG. A total amount of 87.7 trillion angular unknowns was involved in this simulation and required a peak total of 566$GB$ of memory.

\begin{figure}[H]
	\centering
	\includegraphics[width=0.85\linewidth]{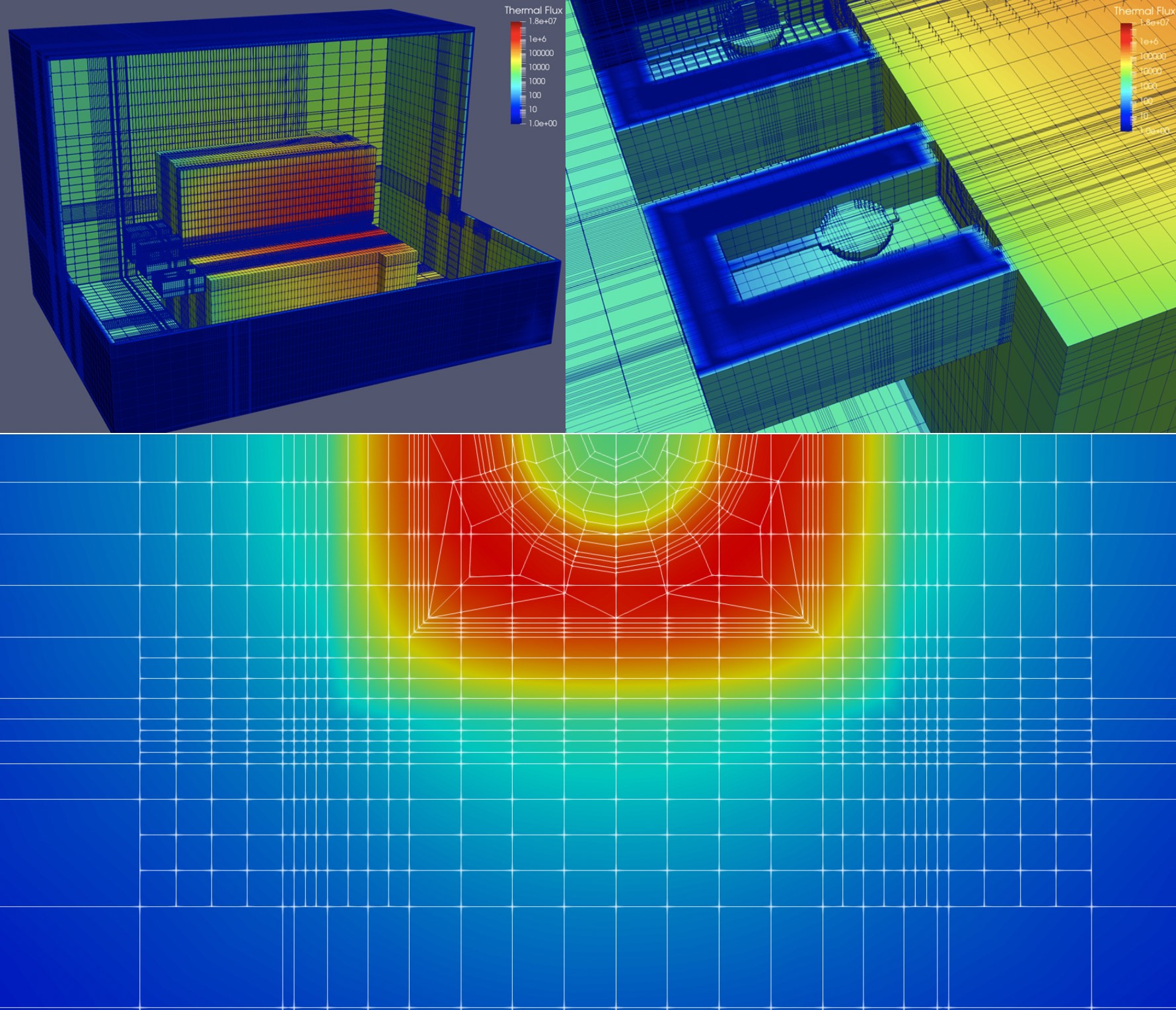}
	\caption{Visualization of the thermal flux of a 116-group large-scale neutron transport simulation. The simulation does not apply ``hanging" nodes but uses degenerate polyhedrons.}
	\label{fig:samplesim}
\end{figure}

\section{Parallel performance - Weak scaling study}
In order to test the sweep machinery at high core counts we performed a weak-scaling study where we constructed orthogonal meshes such that each process has 256 cells per process regardless the number of processes. We used angular quadratures with 3,584 angles total (32 azimuthal angles per octant, 14 polar angles per octant) and varied the amount of groups per groupset to ascertain group aggregation performance. No angle aggregation was applied.

Results of the weak scaling study on LLNL's Quartz machine is shown in Figure \ref{fig:weakscaling} below. The results are for a group aggregation of 64 groups and are shown with Adam's performance model applied relative to the 3,200 core simulations. The reason for applying it relative to 3,200 cores is because the core counts less than $\simeq$2,000 are distributed among a small amount of nodes (44 nodes for 1,568 cores) and it was the author's opinion that these simulations experienced too little across-node communication to justify the chosen communication characteristics. The results show that Chi-Tech operates at $\simeq$84\% parallel efficiency at $\simeq$100k processes and consistently has higher parallel performance than PDT.

In terms of time-to-solution, sample grind times (nanoseconds, $ns$, per angular unknown) are shown in Table 1 below. The grind time results indicate that the group aggregation improves efficiency in both Chi-Tech and PDT, with Chi-Tech taking slightly more advantage. This can be attributed to the factors discussed previously. Primarily Chi-Tech sees more benefit from group aggregation because its data structures are arranged to be contiguous in energy group. It should also be noted that these results are not without a degree of uncertainty because we found the system latency to have some variation in-between runs. If we classify the logarithmic number of processes into small-, medium- and large-amounts then the relative uncertainty was approximately $2\%$, $3\%$ and $4\%$, respectively (according to our initial investigations).
\begin{center}
\begin{minipage}{0.6\textwidth}
	\begin{table}[H]
		\centering
		\label{table:WeakScaling}
		\caption{Grind-times (sweep time per angular unknown) in nanoseconds, $ns$, for mid-scale to large-scale number of parallel processes. Both configurations are compared against varying number of groups}
		{
			\small
		\begin{tabular}{|c|c|c|c|}
			\hline
			
			\textbf{\begin{tabular}[c]{@{}c@{}}Number of \\ processes\end{tabular}} & \textbf{\begin{tabular}[c]{@{}c@{}}Number of \\ groups\end{tabular}} & \multicolumn{1}{l|}{\textbf{\begin{tabular}[c]{@{}l@{}}Grind-time\\ Chi-Tech\\ {[}ns/ang.unk.{]}\end{tabular}}} & \multicolumn{1}{l|}{\textbf{\begin{tabular}[c]{@{}l@{}}Grind-time\\ PDT\\ {[}ns/ang.unk.{]}\end{tabular}}} \\ \hline
			3,200                                                                   & 10                                                                   & 206                                                                                                             & 212                                                                                                        \\ \hline
			3,200                                                                   & 20                                                                   & 158                                                                                                             & 162                                                                                                        \\ \hline
			3,200                                                                   & 64                                                                   & 117                                                                                                             & 154                                                                                                        \\ \hline
			103,968                                                                 & 10                                                                   & 242                                                                                                             & 248                                                                                                        \\ \hline
			103,968                                                                 & 20                                                                   & 179                                                                                                             & 188                                                                                                        \\ \hline
			103,968                                                                 & 64                                                                   & 167                                                                                                             & 181                                                                                                        \\ \hline
		\end{tabular}
	}
	\end{table}
\end{minipage}
\end{center}

\begin{figure} [H]
	\centering
	\includegraphics[width=0.55\linewidth]{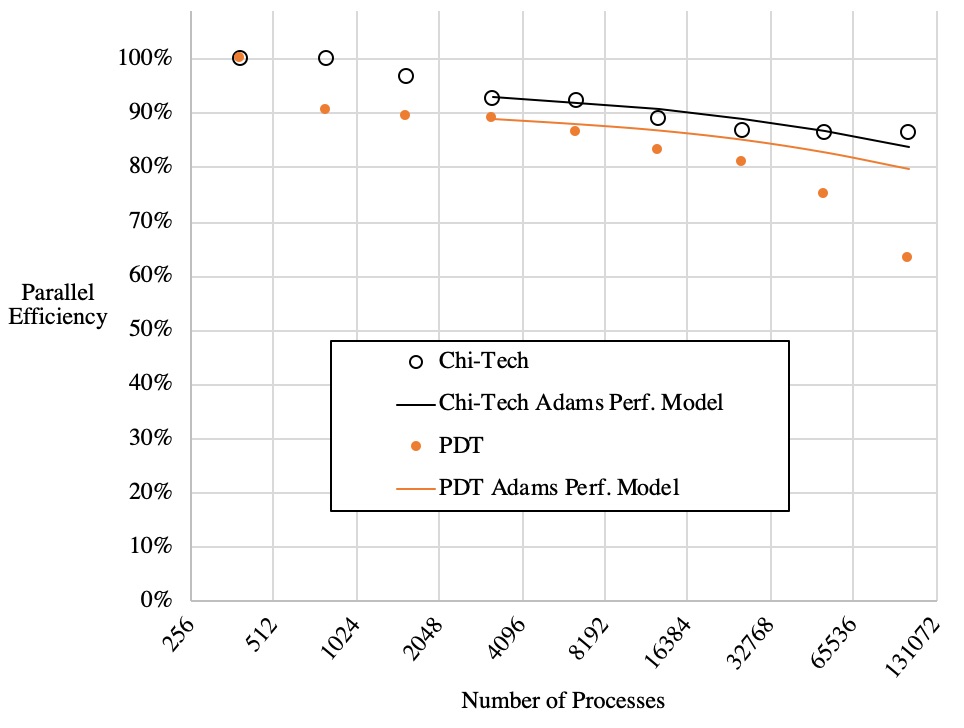}
	\caption{Weak scaling study on LLNL Quartz machine. The results are for 64 energy groups.}
	\label{fig:weakscaling}
\end{figure}

\section{Strong scaling study - with cycles}
\begin{figure}[H]
	\centering
	\includegraphics[width=1.0\linewidth]{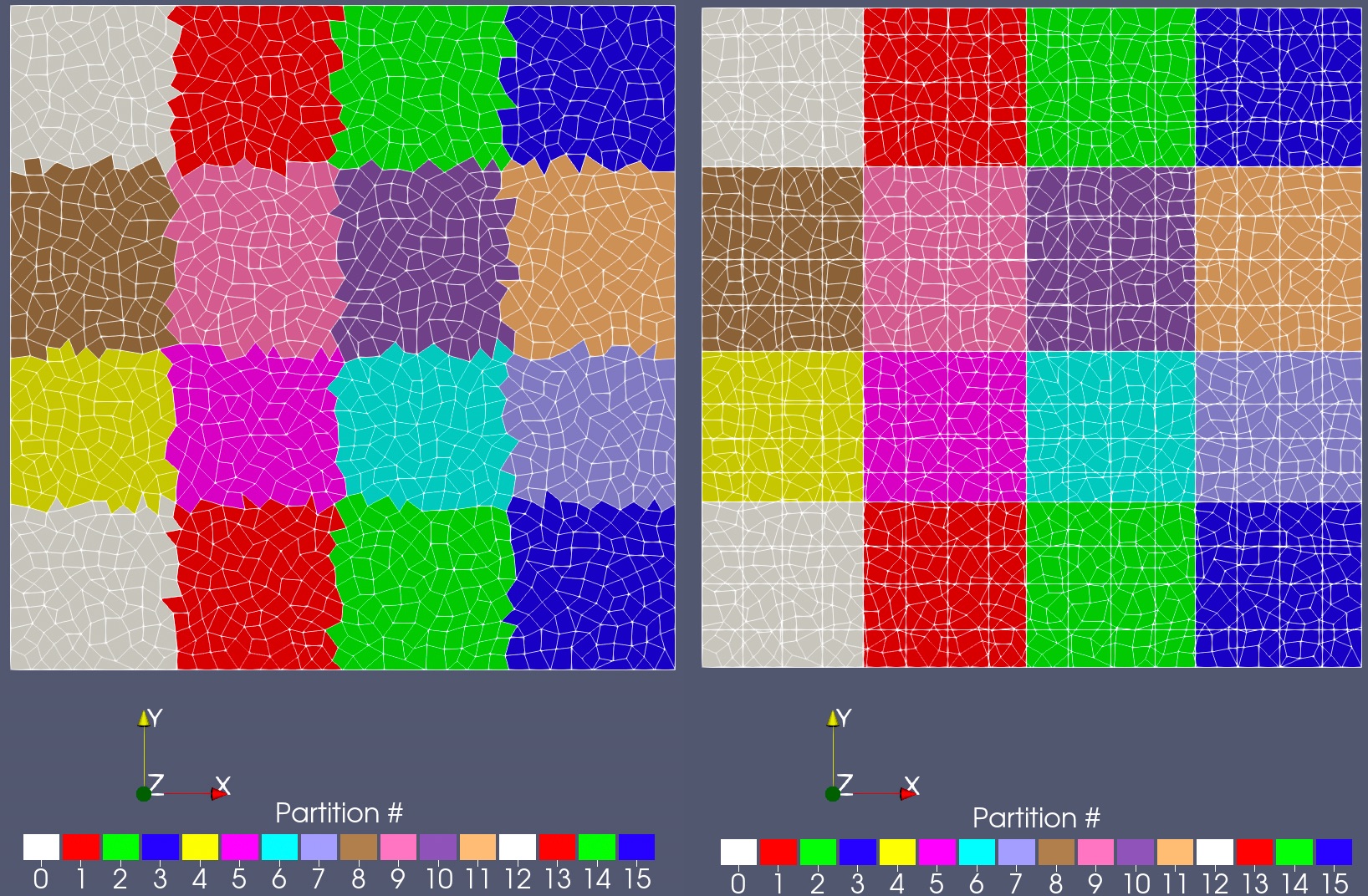}
	\caption{KBA-style partitioning with no cut lines being imprinted (left) versus imprinted (right).}
	\label{fig:cutvsuncutpartitioning}
\end{figure}

With the machinery developed for global cyclic dependencies in Chi-Tech we performed a strong scaling study on two meshes of the same problem shown in Figure \ref{fig:cutvsuncut} below. The problem setup is a block of graphite with center material (shown in red) containing an isotropic source for the first group. Both meshes are axially extruded with the mesh on the left not ``imprinted" with cuts as opposed to the mesh on the right which has imprinted cut-lines in preparation for KBA-style partitioning. The uncut-mesh contains $\sim$72,000 cells where as the cut-mesh contains approximately double that, $\sim$150,000 (extra cells are due to the imprint process).
\begin{figure}[H]
	\centering
	\includegraphics[width=1.0\linewidth]{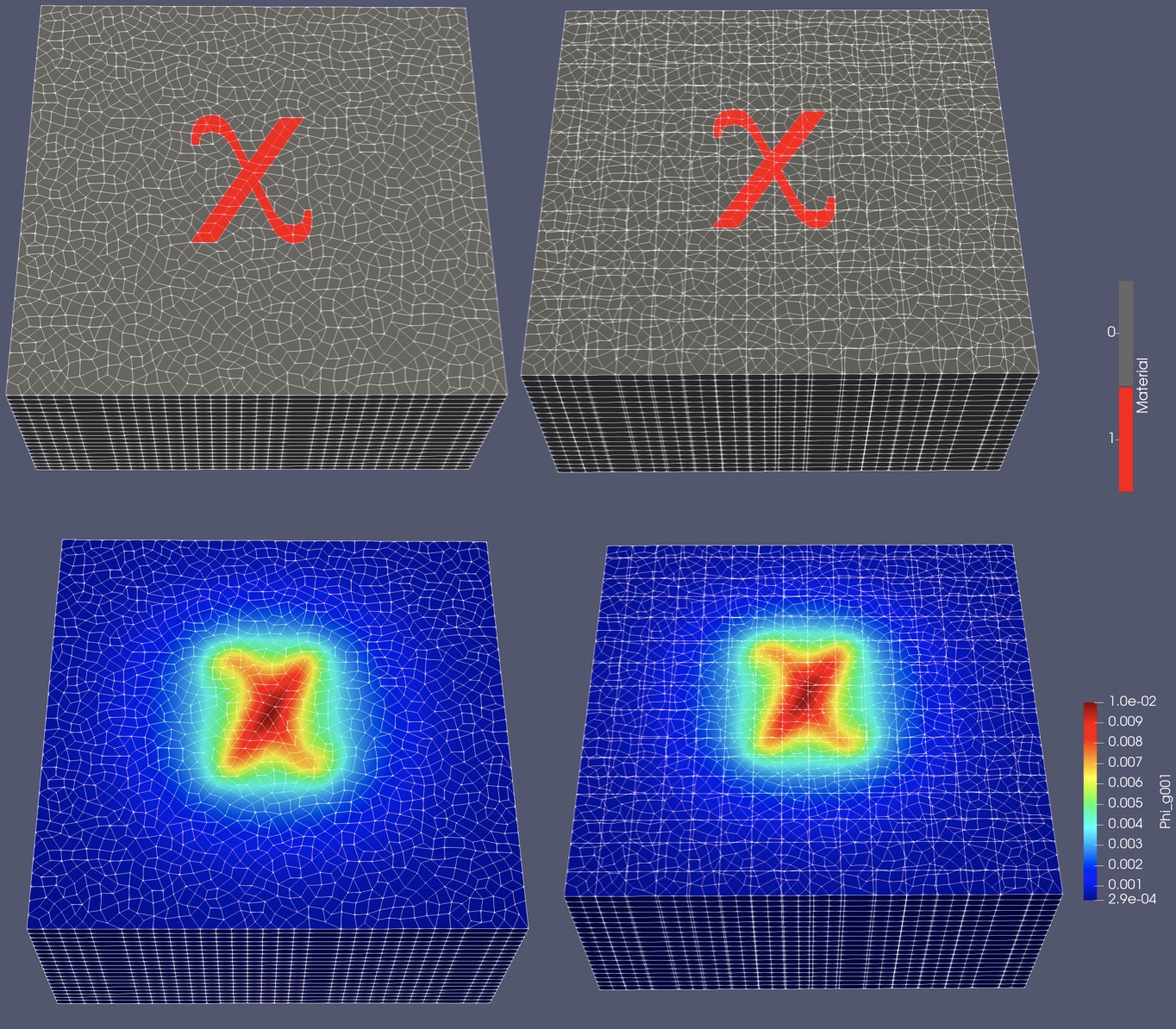}
	\caption{Comparison of a ``cut" versus ``uncut" mesh. The center $\chi$ is the source material (graphite with a source of 1.0 neutrons.cm$^{-3}$ for $g$=1) and the rest is graphite.}
	\label{fig:cutvsuncut}
\end{figure}
A sample partitioning with $P_x {\times}P_y = 4{\times}4$ is shown in Figure \ref{fig:cutvsuncutpartitioning}. This is not the partitioning scheme used for the largest process-count simulation which has too much detail to visualize with colors. We provide this relatively low-process count visualization to aid the reader's understanding. Without cutting the mesh the jagged nature of the partition interfaces can clearly be observed. These jagged faces result in re-entrant conditions and therefore global cyclic dependencies. These dependencies are not present in the cut-mesh. Note also that even though we introduced cycles between partitions, the local cells (i.e., meshes within a given partition) still maintain  their cycle-free character.

\begin{figure}[H]
	\centering
	\includegraphics[width=0.6\linewidth]{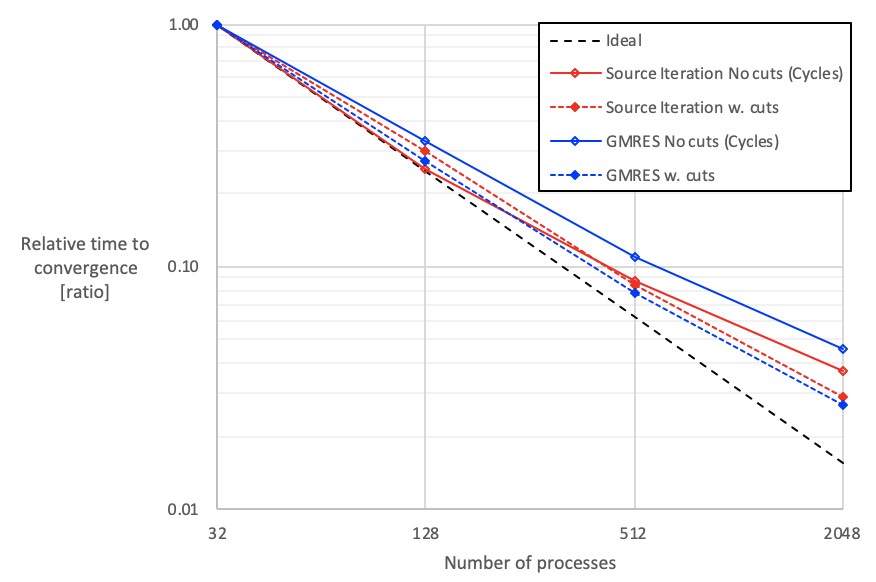}
	\caption{Time-to-solution relative to 32 processes for the uncut- and cut-mesh from a strong scaling study performed on LLNL's Quartz machine.}
	\label{fig:timetosol}
\end{figure}

Relative time-to-solution (not including initialization overheads) for the two mesh types are shown in Figure \ref{fig:timetosol} and the tabulated data is shown in Table \ref{tbl:timetosoldata}. Both the meshes had the same load balancing factor of 1.7. The ratio shown in Figure \ref{fig:timetosol} is taken with respect to a 32 process run.
The uncut mesh shows an increase time-to-solution that is increasing with process count. This can be attributed to two phenomena. The first is the rapid change in cell aggregation, whereby each process receives less cells on which to executes it tasks and therefore altering the time ratio of communication-to-execution time (see Eq.~\eqref{E_opt}), and the second is the increase in idle time incurred by the increase in $P_x$ and $P_y$. 

In Table \ref{tbl:timetosoldata} we observe the introduction of cyclically dependent angular unknowns by the uncut-mesh which form 1.3\% of the angular unknowns, for the 32 process case, and up to 8.3\% for the 2,048 process case. The introduction of these angular fluxes alter the convergence behavior of the iterative methods. For SI (Source Iteration or Richardson iteration), the number of iterations for convergence increased from 12 to 27 (32 processes) by introducing cyclic dependent unknowns, and then increased to 37 from the 32-process run to the 2,048-process run. GMRES showed a similar trend. The cut-mesh shows a total time-to-solution that is 58\% of the ideal scaling at 2,048 process case whereas the uncut-mesh shows 34\%. The additional increase in time-to-solution is primarily caused by the cyclically dependent angular fluxes.

This data represents our first strong scaling study with global cyclic dependencies so we anticipate to improve the efficiency of the algorithm in subsequent studies. Another item of interest is the change in iterative behavior, which we need to investigate and characterize.

This strong-scaling study shows that Chi-Tech can be used to perform large-scale transport simulations on fixed meshes. Where a pure transport simulation is required, the cutline imprint option would still be the best option because it prevents global cyclic dependencies, however, for multi-physics and time dependent simulations it may be beneficial if the transport simulations did not mandate mesh modification. Therefore, the ability to perform transport sweeps on unstructured meshes using volumetric partitioning, as we demonstrated here, is considered to be essential. We demonstrated this ability with Chi-Tech and identified many aspects to be studied. For example, we need to investigate the effect of ParMETIS-type partitioning on parallel sweeps.

\begin{table}[H]
	\centering
	\caption{Time to solution data for a strong scaling study (20 energy groups each) using 2048 angles performed on LLNL's Quartz machine that compares the uncut-mesh to the cut-mesh described above.}
	\includegraphics[width=0.9\linewidth]{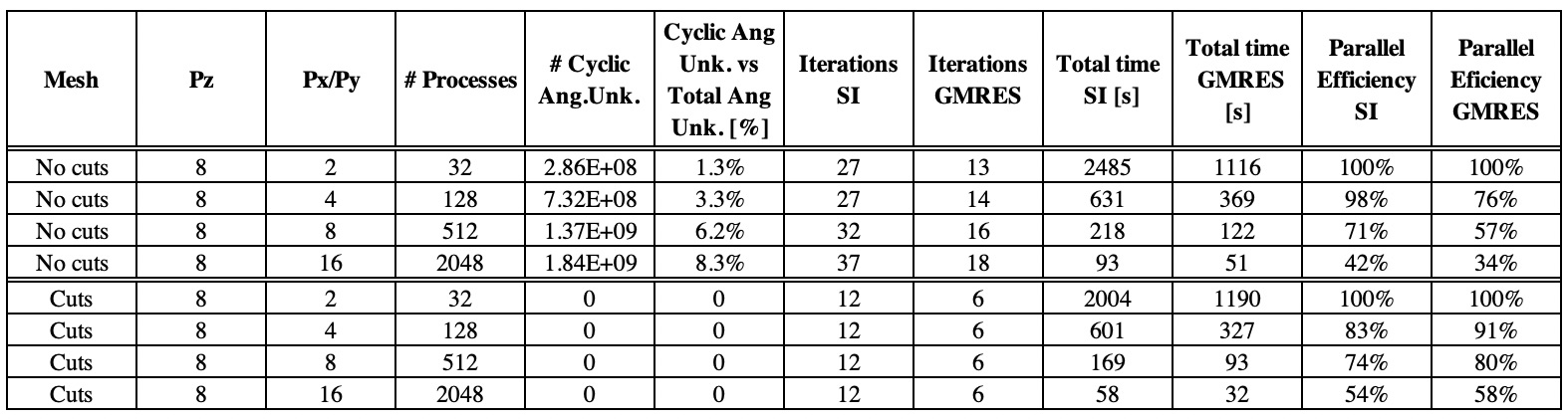}
	\label{tbl:timetosoldata}
\end{table}

\section{Conclusions and Outlook}
We have developed a new Sweep Plane Data Structure (SPDS) that allows for massively parallel transport sweeps with cyclic dependencies. This data structures functions well within our new simulation platform Chi-Tech which offers exciting opportunities with regards to unstructured meshes and multi-physics simulations. In particular, we have demonstrated that our implementation follows the performance model in \cite{ProvablyOptimal}. 

We demonstrated the code's ability to perform realistic, non-reactor, transport problems and showed that these type of simulations can be performed efficiently and in a timely manner given the appropriate computing resources.
Our weak-scaling study shows that Chi-Tech can be used to solve exascale simulations on modern computing architectures as was demonstrated with simulations using over 100,000 processes.
We also demonstrated the codes ability to handle volumetric partitioning schemes that will be conducive to multi-physics simulations.

In future work we plan to investigate the performance related aspects of using partitioning schemes common to other physics simulations (i.e., ParMETIS) as well as to extend our mesh infrastructure, and our SPDS, to high-order meshes.

\newpage
\section{Acknowledgments}
The fundamental mechanisms of our SPDS stemmed from the work by Marvin Adams, Michael Adams, and Daryl Hawkins. We extend our gratitude for many fruitful discussions.

The computing resources available in this work was made possible by involvement in Texas A\&M's Center for Exascale Radiation Transport which was supported by the Department of Energy, National Nuclear
Security Administration, under Award Number(s) DE-NA0002376. Established by Congress in 2000, NNSA is a semi-autonomous agency within the U.S. Department of Energy responsible for enhancing national security through the military application of nuclear science. NNSA maintains and enhances the safety, security, reliability and performance of the U.S. nuclear weapons stockpile without nuclear testing; works to reduce global danger from weapons of mass destruction; provides the U.S. Navy with safe and effective nuclear propulsion; and responds to nuclear and radiological emergencies in the U.S. and abroad.
One of the authors (JR) was partially funded through a grant by the Department of the Defense, Defense Threat Reduction Agency under Award No. HDTRA1-18-1-0020. The content of the information does not necessarily reflect the position or the policy of the federal government, and no official endorsement should be inferred.


\begin{thebibliography}{1}
	
	\bibitem{LewisMiller} Lewis E.E., Miller W.F., {\em Computational Methods of Neutron Transport}, JohnWiley \& Sons, 1984
	
	\bibitem{KBA} Randal S. Baker \& Kenneth R. Koch (1998) An Sn Algorithm for the Massively
	Parallel CM-200 Computer, Nuclear Science and Engineering, 128:3, 312-320, DOI: 10.13182/
	NSE98-1
	
	 \bibitem{ProvablyOptimal0} Adams et al., {\em Provable optimal parallel transport sweeps on regular grids}, International Conference on Mathematics and Computational Methods Applied to Nuclear Science and Engineering (M\&C). 2013
	 
	 \bibitem{ProvablyOptimalb} Adams et al.,  “Provably Optimal Parallel Transport
	 Sweeps with Non-Contiguous Partitions,” in “ANS
	 MC2015-Joint International Conference on Mathematics
	 and Computation (M\&C), Supercomputing in Nuclear Applications (SNA) and the Monte Carlo (MC) Method", 2015
	 
	 \bibitem{Analysis2009} T. S. Bailey and R. D. Falgout, “Analysis Of Massively Parallel Discrete-Ordinates Transport Sweep Algorithms With Collisions,” Proc. International Conference on Mathematics, Computational Methods \& Reactor Physics, Saratoga Springs, May 3-7, CDROM (2009).
	
	 \bibitem{ProvablyOptimal} Adams et al., {\em Provable optimal parallel transport sweeps on semi-structured grids}, Journal of Computational Physics, DOI: 10.1016/j.jcp.2020.109234, January 2020.
	 
	 \bibitem{STAPL} G. Tanase, A. Buss, A. Fidel, Harshvardan, I. Papadopoulos, O. Pearce, T. Smith, N. Thomas, X. Su, N. Mourad, J. Vu, M. Bianco, N. M. Amato, L. Rauchwerger, “The STAPL Parallel Container Framework,” Proc. ACM SIGPLAN Symp. Prin. Prac. Par. Prog. (PPOPP), (2011).
	
	\bibitem{Stone} H. G. Stone and M. L. Adams, “A Piecewise Linear Finite Element Basis with Application to Particle Transport,” Proc. International Conf. on Nuclear Mathematical and Computational Sciences, Gatlin- burg, TN, April 6-10 (2003).
	
	\bibitem{Ghaddar2017} Ghaddar T.H., Ragusa J.C., {\em An approach for Load Balancing Massively Parallel Transport Sweeps on Unstructured Grids}, International Conference on Mathematics and Computational Methods Applied to Nuclear Science and Engineering, April 2017.
	
	\bibitem{Ghaddar2019} Ghaddar T., Ragusa J.C., Vermaak J.I.C., “Approaches to Load Balancing Massively Parallel Transport Sweeps on Unstructured Grids”, Joint International Conference on Mathematics and Computation, 2019
	
	\bibitem{ParMETIS} Karypis G. {\em ParMETIS - Parallel Graph Partitioning and Fill-reducing Matrix Ordering}, http://glaros.dtc.umn.edu/gkhome/metis/parmetis/overview, date accessed January 30, 2020
	
	\bibitem{PETSc} Satish Balay and Shrirang Abhyankar and Mark~F. Adams and Jed Brown and Peter Brune
	and Kris Buschelman and Lisandro Dalcin and Alp Dener and Victor Eijkhout and William~D. Gropp
	and Dmitry Karpeyev and Dinesh Kaushik and Matthew~G. Knepley and Dave~A. May and Lois Curfman McInnes
	and Richard Tran Mills and Todd Munson and Karl Rupp and Patrick Sanan
	and Barry~F. Smith and Stefano Zampini and Hong Zhang and Hong Zhang, {\em PETSc Users manual}, Argonne National Laboratory, ANL-95/11 Revision 3.12, 2019
	
	\bibitem{HYPRE} {\em User manual: HYPRE - High Performance Preconditioners}, Lawrence Livermore National Laboratory - Center for Applied Scientific Computing, Version 2.11.1
	
	
	
	\bibitem{Abu1} Abu-Shumays I., {\em Compatible Product Angular Quadrature for Neutron Transport in X-Y Geometry}. Nuclear Science and Engineering 64 (1977), 299–316.
	
	\bibitem{Abu2} Abu-Shumays I., {\em Angular Quadratures for Improved Transport Computations}, Transport Theory and Statistical Physics 30(2\&3) (2001), 169–204.
	
	\bibitem{Abu3} Abu-Shumays I., Yehnert C., {\em Angular Interpolations and Splice Options for Three- Dimensional Transport Computations} Tech. Rep. DE-AC11-93PN38195, Bettis Atomic Power Laboratory, Pittsburgh, PA, 1996.
	
	\bibitem{LDFE} Jarrel J.J., Adams M.L., {\em Discrete-Ordinates Quadrature sets based on Linear Discontinuous Finite Elements}, International conference on Mathematics and Computational Methods applied to Nuclear Science and Engineering, May 2011.
	
	\bibitem{LDFE2} Cheuk Y. Lau \& Marvin L. Adams (2017) Discrete Ordinates Quadratures
	Based on Linear and Quadratic Discontinuous Finite Elements over Spherical Quadrilaterals,
	Nuclear Science and Engineering, 185:1, 36-52, DOI: 10.13182/NSE16-28
    
    \bibitem{GMRESSaad} Saad Y., Schultz M.H., {\em GMRES: A Generalized Minimal Residual algorithm for solving non-symmetric linear systems}, SIAM Journal of Scientific and Statistical Computation, volume 7, pages 856-869, 1986.
    
    \bibitem{GMRES0} Oliveira S., Deng Y., {\em Preconditioned Krylov Subspace Methods for Transport Equations}, Progress in Nuclear Energy 33, 155. 1998
    
    \bibitem{GMRES1} Guthrie B., Holloway J.P., and Patton B.W., {\em GMRES as a Multi-Step Transport Sweep Ac- celerator}, Transport Theory Statist. Phys. 28, 83. 1999
    
    \bibitem{GMRES2} Patton B.W., Holloway J.P., {\em Application of preconditioned GMRES to the numerical solution of the neutron transport equation}, Annals of Nuclear Energy volume 29, issue 2, Pages 109-136, January 2002.
    
    \bibitem{Adamslarsen} M. L. Adams and E. W. Larsen, ``Fast iterative methods for discrete-ordinates particle transport calculations,” Prog. Nucl. Energy, 40, No. 1, pp. 3-159, (2002).
    
    \bibitem{PolyFEM} Bailey T., Adams M.L., {\em The Piecewise Linear Discontinuous Finite Element Method applied to the RZ and XYZ Transport Equations}, PhD Dissertation, Texas A\&M University, May 2008.
    
    \bibitem{Poly2FEM} Bailey et al., {\em A piecewise linear finite element discretization of the diffusion equation for arbitrary polyhedral grids}, Journal of Computational Physics vol. 227, pages 3738-3757, March 2008.
    
    \bibitem{TopologicalSorting} Cormen, Thomas H.; Leiserson, Charles E.; Rivest, Ronald L.; Stein, Clifford (2001), {\em Section 22.4: Topological sort, Introduction to Algorithms (2nd ed.)}, MIT Press and McGraw-Hill, pp. 549–552, ISBN 0-262-03293-7.
    
    \bibitem{ValidationSweeps} Bailey et al., {\em Validation of Full-Domain Massively Parallel Transport Sweep Algorithms}
    
   
    
    \bibitem{WIMS} Askew J.R., Fayers F.J., Kemshell P.B.,  {\em A general
    description of the lattice code WIMS}. Journal of the British
    Nuclear Energy Society (October), 564. 1996.
    
    \bibitem{CurvedGrids} Liu C., Larsen E.W., {\em Discrete Ordinates Methods for Transport Problems with Curved Spatial Grids}, PhD dissertation, University of Michigan, Department of Nuclear Engineering and Radiological Sciences, 2015.
    
    \bibitem{EdgeSplitting} T. S. Haut, P. G. Maginot, V. Z. Tomov, B. S. Southworth, T. A. Brunner \& T. S. Bailey (2019) {\em An Efficient Sweep-Based Solver for the SN Equations on High-Order Meshes}, Nuclear Science and Engineering, 193:7, 746-759, DOI: 10.1080/00295639.2018.1562778
    
    \bibitem{RemoveCycles} Steven J. Plimpton, Bruce Hendrickson, Shawn P. Burns, William McLendon III \& Lawrence Rauchwerger (2005) Parallel Sn Sweeps on Unstructured Grids: Algorithms for Prioritization, Grid Partitioning, and Cycle Detection, Nuclear Science and Engineering, 150:3, 267-283, DOI: 10.13182/NSE150-267
    
    \bibitem{RemoveCycleLLNL} T. S. Haut, P. G. Maginot, V. Z. Tomov, T. A. Brunner, T. S. Bailey, {\em An Efficient Sweep-based Solver for the Sn Equations on High-Order Meshes}, Lawrence Livermore National Laboratory Document, LLNL-PROC-744912, January 2018.
    
    \bibitem{RemoveCyclesYaqi} Yaqi Wang \& Jean C. Ragusa (2010) Diffusion Synthetic Acceleration for High-Order Discontinuous Finite Element SN Transport Schemes and Application to Locally Refined Unstructured Meshes, Nuclear Science and Engineering, 166:2, 145-166, DOI: 10.13182/ NSE09-46
    
    \bibitem{MIP} Turcksin, Bruno, and Ragusa, Jean C., "Discontinuous diffusion synthetic acceleration for S{sub n} transport on 2D arbitrary polygonal meshes". United States: N. p., 2014. Web. doi:10.1016/J.JCP.2014.05.044.
    
    \bibitem{WGDSA} Edward W. Larsen (1984) Diffusion-synthetic acceleration methods for discrete-ordinates problems, Transport Theory and Statistical Physics, 13:1-2, 107-126, DOI: 10.1080/00411458408211656
    
    \bibitem{WGDSA2} Marvin L. Adams \& William R. Martin (1992) Diffusion Synthetic Acceleration
    of Discontinuous Finite Element Transport Iterations, Nuclear Science and Engineering, 111:2,
    145-167, DOI: 10.13182/NSE92-A23930
    
    \bibitem{TwoGrid} B. T. Adams \& J. E. Morel (1993) A Two-Grid Acceleration Scheme for the
    Multigroup Sn Equations with Neutron Upscattering, Nuclear Science and Engineering, 115:3,
    253-264, DOI: 10.13182/NSE115-253
    
    \bibitem{ChiTech} Vermaak J.I.C., {\em $\chi$-Tech: A large-scale scientific simulation engine being developed at Texas A\&M University as part of a research study}, {https://github.com/chi-tech/chi-tech}
    
    \bibitem{SERPENT} Leppänen, J., et al. (2015) "The Serpent Monte Carlo code: Status, development and applications in 2013." Ann. Nucl. Energy, 82 (2015) 142-150.
    
    
    
    \bibitem{CERT} {https://cert.tees.tamu.edu/}
    
    
\end{thebibliography}
\end{document}